\documentclass[twocolumn,pra,showpacs,superscriptaddress]{revtex4-1}
\pdfoutput=1
\usepackage{setspace}
\usepackage{amsfonts,amsmath,amssymb}
\usepackage{txfonts}
\usepackage{graphicx, color}
\usepackage{bm}

\newcommand{\reff}[1]{(\ref{#1})}
\begin{document}

\title{Many-body physics in the classical-field description of a degenerate Bose gas}

\author{T. M. Wright}
\affiliation{The University of Queensland, School of Mathematics and Physics, ARC Centre of Excellence for Quantum-Atom Optics, Qld 4072, Australia}
\author{N. P. Proukakis}
\affiliation{School of Mathematics and Statistics, Newcastle University, Newcastle upon Tyne, NE1 7RU, United Kingdom}
\affiliation{The University of Queensland, School of Mathematics and Physics, ARC Centre of Excellence for Quantum-Atom Optics, Qld 4072, Australia}
\author{M. J. Davis}
\affiliation{The University of Queensland, School of Mathematics and Physics, ARC Centre of Excellence for Quantum-Atom Optics, Qld 4072, Australia}

\begin{abstract}
The classical-field formalism has been widely applied in the calculation of normal correlation functions, and the characterization of condensation, in finite-temperature Bose gases.  Here we discuss the extension of this method to the calculation of more general correlations, including the so-called anomalous correlations of the field, without recourse to symmetry-breaking assumptions.  Our method is based on the introduction of $\mathrm{U(1)}$-symmetric classical-field variables analogous to the modified quantum ladder operators of number-conserving approaches to the degenerate Bose gas, and allows us to rigorously quantify the anomalous and non-Gaussian character of the field fluctuations.  We compare our results for anomalous correlation functions with the predictions of mean-field theories, and demonstrate that the nonlinear classical-field dynamics incorporate a full description of many-body processes which modify the effective mean-field potentials experienced by condensate and noncondensate atoms.  We discuss the role of these processes in shaping the condensate mode, and thereby demonstrate the consistency of the Penrose-Onsager definition of the condensate orbital in the classical-field equilibrium.  We consider the contribution of various noncondensate-field correlations to the overall suppression of density fluctuations and interactions in the field, and demonstrate the distinct roles of phase and density fluctuations in the transition of the field to the normal phase.      
\end{abstract}


\date{\today}

\maketitle
\section{Introduction}\label{sec:Introduction}
The experimental realization of Bose-Einstein condensation (BEC) in dilute atomic gases~\cite{Anderson95,Davis95,Bradley95} has lead to a resurgence of interest in the theory of weakly interacting Bose gases~\cite{Shi98,Dalfovo99,Proukakis08}.  Theories of weakly-interacting BEC were initially developed~\cite{Bogoliubov47,Beliaev58} in the hope of obtaining insight into the physics of the strongly interacting superfluid helium, and can only give a qualitative account of the physics of the liquid superfluid phase~\cite{Griffin93}.  By contrast, the advent of dilute, weakly interacting Bose gases in the laboratory provides for the direct comparison of theories of weakly interacting BEC with experiments (for a review see Ref.~\cite{Proukakis08}).  More generally, these systems offer an unprecedented opportunity for the quantitative experimental evaluation of quantum-field models at finite temperatures, and away from equilibrium, as noted by other authors~\cite{Stoof99,Rusch00,Morgan03,Morgan04}.  

The analysis of the weakly interacting Bose gas at finite temperature is significantly complicated by the necessity of considering interactions between excitations of the condensate.  A self-consistent mean-field approach, based on approximate factorizations of field-operator products~\cite{Griffin96} leads to the so-called Hartree-Fock-Bogoliubov (HFB) formalism for the finite-temperature field, which is also obtained in a variational approach to the problem~\cite{Blaizot85,Proukakis96}.  This description, however, violates known exact constraints on the excitation spectrum~\cite{Hugenholtz59}, due to the inconsistent introduction of (leading-order) many-body effects~\cite{Bijlsma97} into the description of atomic interactions.  The simplest resolution of this problem is provided by the so-called Popov approximation~\cite{Shi94,Griffin96} to the HFB theory (HFB-Popov), in which interactions between excitations at the Hartree-Fock level are retained~\cite{Shi98}, while many-body corrections are neglected in the treatment of all interatomic interactions. 

A systematic perturbative treatment of the beyond-quadratic terms of the Bose-field Hamiltonian~\cite{Morgan00} (see also Refs.~\cite{Fedichev98b,Shi98}) demonstrates that the failure of the HFB method results from an inconsistent treatment of Hamiltonian terms cubic in the Bose field operator in the mean-field factorization approximation.  This provides support for an extension of the conventional HFB approach~\cite{Proukakis98b,Hutchinson00} in which the Popov theory is augmented by spatially dependent \emph{effective} interaction strengths, which serve to upgrade the Hartree-Fock interactions between atoms to interactions mediated by a many-body $T$~matrix, which accounts for all ladder-diagram processes contributing to interactions in the finite-temperature medium~\cite{Stoof96,Bijlsma97,Shi98,Proukakis98a}.  However, in this theory the zero-energy limit of the $T$~matrix is substituted for the interaction strength in all condensate-excitation interactions, neglecting the dependence of the $T$~matrix on the collision energies.  More complicated (bubble-diagram~\cite{Bijlsma96}) effects which appear at the same order of perturbation theory~\cite{Morgan00} as the ladder-diagram corrections are also neglected.  Kinetic theories which include processes beyond those contained in the HFB theories, such as the exchange of atoms between the condensate and thermal cloud, have been considered by several authors~\cite{Proukakis96,Walser99,Walser00,Giorgini00,Wachter01,Proukakis01,Imamovic01}.

An alternative approach to modeling the finite-temperature Bose gas is provided by the so-called classical-field (or c-field) formalism~\cite{Blakie08,Kagan97,Lobo04,Brewczyk07}.  In this approach, one considers the dynamics of a classical dynamical system corresponding to the high mode-occupation limit of the formal second-quantized field theory~\cite{Fetter71a,Blaizot85} for the dilute Bose gas.  The classical-field model provides a leading-order description of the long-wavelength physics which dominate the critical behavior associated with the Bose-condensation phase transition~\cite{Baym99}.  More generally, the \emph{dynamical} classical-field equations of motion arise as the ``classical'' component of the atomic-field evolution in the Wigner representation of the second-quantized field~\cite{Steel98,Sinatra01,Polkovnikov03,Norrie05}.  The classical-field model is thus expected to give a good description of the most highly occupied modes of the system, including the condensate and its low-lying excitations, for which quantum fluctuations can be safely neglected.  The introduction of an explicit single-particle energy cutoff restricts the system to these low-energy modes~\cite{Davis01,Blakie05}.  In closely related stochastic-field methods~\cite{Stoof99,Gardiner03}, the description of the low-energy modes includes explicit damping and noise terms arising from the coupling of these modes to the eliminated high-energy component of the Bose field.  The classical wave-mixing dynamics of the resulting (Hamiltonian or stochastic) field equation of motion then provide an intrinsic nonperturbative description of all interaction processes in the low-energy field, neglecting only the quantal nature of excitations, which is essentially irrelevant in the high-temperature regime.  The inclusion of many-body effects beyond Bogoliubov theory in the Hamiltonian classical-field theory, and agreement of this theory with the second-order perturbative treatment of Ref.~\cite{Morgan00} was demonstrated for a homogeneous field in Refs.~\cite{Davis01,Davis02}. 

In this article, we present a comprehensive, explicit demonstration that the equilibrium classical-field dynamics provide an intrinsic description of many-body interactions in the finite-temperature, harmonically confined Bose field.  Introducing appropriate $\mathrm{U(1)}$-symmetric field variables, analogous to the modified ladder operators of number-conserving Bogoliubov theories~\cite{Girardeau59,Gardiner97a,Castin98,Girardeau98}, we quantify the anomalous and non-Gaussian nature of the field fluctuations.  These correlations reveal signatures of many-body processes neglected in (e.g.) the HFB theories.  We discuss the importance of these processes in shaping the condensate orbital, and demonstrate the consistency of the Penrose-Onsager definition of condensation in the classical-field equilibrium.  In particular, we show that the Penrose-Onsager condensate appears as an effective eigenfunction of the total mean-field potential it experiences, provided that the contributions of anomalous averages (both pair and triplet) to the latter are taken into account.  We also consider the temperature dependence of the many-body effects, and discuss the relation between condensation and the overall suppression of density fluctuations in the system.  Our results reveal that the partially condensed Bose gas exhibits non-trivial correlations as a result of purely thermal (classical) fluctuations of the atomic field, which dominate over the quantum fluctuations of the field in realistic experimental systems~\cite{NoteA}. 

This article is organized as follows: In Sec.~\ref{sec:Formalism} we briefly review the derivation of the projected Gross-Pitaevskii equation (PGPE) formalism, and give the parameters of the system we investigate.  In Sec.~\ref{sec:Correlation_functions} we describe the microcanonical interpretation of the PGPE, and define $\mathrm{U(1)}$-symmetric classical-field variables appropriate for the evaluation of anomalous correlation functions of the classical field.  In Sec.~\ref{sec:Correlations} we discuss the local correlation functions of the noncondensed component of the field and their relation to many-body interaction effects described by the classical-field model.  In Sec.~\ref{sec:E_dependence} we consider the dependence of the noncondensate correlations and interaction effects on the energy (and thus temperature) of the classical-field equilibrium. In Sec.~\ref{sec:Suppression} we consider the various contributions to the suppression of density fluctuations in the field which are neglected in mean-field theories. In Sec.~\ref{sec:Conclusions} we summarize and present our conclusions.  

\section{Formalism}\label{sec:Formalism}
\subsection{PGPE formalism}\label{subsec:PGPE_formalism}
A detailed review of the formalism of (projected) classical-field methods has recently been given in Ref.~\cite{Blakie08}, but for the reader's convenience, we briefly describe the relevant details of the formalism here.

Formally, the physics of the harmonically trapped dilute Bose gas is governed by the second-quantized Hamiltonian
\begin{eqnarray}\label{eq:H_fundamental}
    \hat{H} &=& \int\!d\mathbf{x}\,\hat{\Psi}^\dagger(\mathbf{x})H_\mathrm{sp}\hat{\Psi}(\mathbf{x}) \nonumber \\
       &&+ \frac{1}{2}\int\!d\mathbf{x}\!\int\!d\mathbf{x}'\,\hat{\Psi}^\dagger(\mathbf{x})\hat{\Psi}^\dagger(\mathbf{x}') U(\mathbf{x}-\mathbf{x}')\hat{\Psi}(\mathbf{x}')\hat{\Psi}(\mathbf{x}), 
\end{eqnarray}
where the single-particle Hamiltonian is
\begin{equation}
	H_\mathrm{sp} = \frac{-\hbar^2\nabla^2}{2m} + \frac{m}{2}\Big[\omega_x^2x^2 + \omega_y^2y^2 + \omega_z^2z^2\Big],
\end{equation}
and $U(\mathbf{x})$ is the exact interatomic potential.  
We introduce a single-particle subspace $\mathbf{L}$ spanned by eigenmodes $Y_n(\mathbf{x})$ of the single-particle Hamiltonian [$H_\mathrm{sp}Y_n(\mathbf{x})=\epsilon_nY_n(\mathbf{x})$] with energies $\epsilon_n$ less than some cutoff $E_\mathrm{max}$, and a complementary subspace comprised of the remaining high-energy modes.  Provided $E_\mathrm{max}$ is chosen such that the high-energy modes are essentially unoccupied, the dynamics of these modes can be integrated out to obtain an effective Hamiltonian for the low-energy (coarse-grained) Bose field $\hat{\Psi}_\mathbf{L}(\mathbf{x})=\sum_{n\in\mathbf{L}}\hat{a}_nY_n(\mathbf{x})$, as shown by Morgan~\cite{Morgan00}.  Atomic interactions described by the effective Hamiltonian are mediated by an approximate 2-body $T$~matrix, and the interaction can thus be rigorously approximated by a ``contact" potential, with a renormalized coupling constant $U_0$.  The low-energy Hamiltonian then takes the form
\begin{align}\label{eq:H_L}
    \hat{H}_\mathrm{L} = \int\!&d\mathbf{x}\,\hat{\Psi}_\mathbf{L}^\dagger(\mathbf{x})H_\mathrm{sp}\hat{\Psi}_\mathbf{L}(\mathbf{x})  \nonumber \\
    &+ \frac{U_0}{2}\!\int\!d\mathbf{x}\,\hat{\Psi}_\mathbf{L}^\dagger(\mathbf{x})\hat{\Psi}_\mathbf{L}^\dagger(\mathbf{x})\hat{\Psi}_\mathbf{L}(\mathbf{x})\hat{\Psi}_\mathbf{L}(\mathbf{x}),  
\end{align}
which defines an effective field theory~\cite{Andersen04} for the coarse-grained field $\hat{\Psi}_\mathbf{L}(\mathbf{x})$.

We then further divide the low-energy region $\mathbf{L}$ into a coherent region (or condensate band) $\mathbf{C}=\{n : \epsilon_n < \epsilon_\mathrm{cut}\}$, spanned by single-particle eigenmodes $Y_\mathrm{n}(\mathbf{x})$ with energies below some classical-field cutoff $\epsilon_\mathrm{cut}$, and a complementary incoherent region $\mathbf{I}=\{n : \epsilon_\mathrm{cut} \leq \epsilon_n < E_\mathrm{max}\}$.  
Introducing the projector
\begin{equation}\label{eq:Pdef} 
    \mathcal{P}\big\{f(\mathbf{x})\big\}\equiv\sum_{n \in  \mathbf{C}}Y_n(\mathbf{x})\!\int\!d\mathbf{y}\, Y_n^*(\mathbf{y})f(\mathbf{y}),  
\end{equation} 
onto the coherent region $\mathbf{C}$, we define a coherent-region field operator
\begin{equation}\label{eq:C_region_opr}
    \hat{\psi}(\mathbf{x}) \equiv  \mathcal{P}\big\{ \hat{\Psi}_\mathbf{L}(\mathbf{x}) \big\} = \sum_{n\in\mathbf{C}} \hat{a}_n Y_n(\mathbf{x}).
\end{equation}
Neglecting the coupling of this field operator to modes of the field in the incoherent region $\mathbf{I}$, we find that $\hat{\psi}(\mathbf{x})$ is governed by a Hamiltonian obtained from Eq.~\reff{eq:H_L} by the replacement $\hat{\Psi}_\mathbf{L}(\mathbf{x})\to\hat{\psi}(\mathbf{x})$.  The classical-field approximation is then made by demoting the operators $\hat{a}_n$ in Eq.~\reff{eq:C_region_opr} to classical variables $\alpha_n$, thereby defining the classical field $\psi(\mathbf{x})\equiv\sum_{n\in\mathbf{C}} \alpha_n Y_n(\mathbf{x})$.  The evolution of the field $\psi(\mathbf{x})$ is governed by the classical-field Hamiltonian
\begin{equation}\label{eq:HCF}
    H_\mathrm{CF}[\psi] = \int\!d\mathbf{x}\,\psi^*(\mathbf{x})H_\mathrm{sp}\psi(\mathbf{x}) + \frac{U_0}{2}|\psi(\mathbf{x})|^4,
\end{equation}
and Hamilton's equations for the classical variables $\alpha_n$ obtained from Eq.~\reff{eq:HCF} can be expressed concisely as the field equation of motion
\begin{equation}\label{eq:PGPE} 
    i\hbar\frac{\partial\psi(\mathbf{x})}{\partial t} = \mathcal{P}\left\{\left(H_{\mathrm{sp}}+U_0|\psi(\mathbf{x})|^2\right)\psi(\mathbf{x})\right\},
\end{equation} 
which is termed the projected Gross-Pitaevskii equation \cite{Blakie08}.  Equation~\reff{eq:PGPE} describes the classical (thermal) fluctuations of the coherent region, while neglecting the contribution of quantum (vacuum) fluctuations, which is a valid approximation in regimes of significant thermal excitation, where mode occupations are large, and thermal fluctuations dominate.  For an appropriately chosen cutoff $\epsilon_\mathrm{cut}$, one can augment the classical-field description of the coherent region $\mathbf{C}$ with a mean-field (Hartree-Fock) description of the incoherent region $\mathbf{I}$, allowing for accurate quantitative predictions for the full Bose-field system~\cite{Davis06}.  In this article, we consider only the correlations of the coherent region $\mathbf{C}$ that arise from the kinetics of Eq.~\reff{eq:PGPE}.
\subsection{System parameters}\label{subsec:System_parameters}
For the purpose of the numerical implementation of the PGPE, we rescale Eq.~\reff{eq:PGPE} by introducing units of distance $r_0=\sqrt{\hbar/m\omega_r}$, time $t_0=\omega_r^{-1}$ and energy $\epsilon_0 = \hbar\omega_r$, with $\omega_r=\omega_x=\omega_y$ the transverse trapping frequency of a cylindrically symmetric system.  With these choices, Eq.~\reff{eq:PGPE} becomes
\begin{equation}
	i\frac{\partial \psi}{\partial t} = -\frac{1}{2} \nabla^2\psi +\frac{1}{2}(x^2 + y^2 + \lambda_z^2z^2)\psi + \mathcal{P}\big\{C_\mathrm{nl}|\psi|^2\psi\big\},
\end{equation}
where we have chosen the classical field $\psi(\mathbf{x})$ to be unit-normalized, and absorbed the total classical-field atom number $N_\mathrm{c}$ into the nonlinear coefficient $C_\mathrm{nl} = N_\mathrm{c}U_0/\hbar\omega_rr_0^3$.  We follow Ref.~\cite{Wright10b} in choosing parameters $\lambda_z=\sqrt{8}$ (a typical three-dimensional trap geometry), $C_\mathrm{nl}=\sqrt{2}\times 500$, and $\epsilon_\mathrm{cut} = 31\hbar\omega_r$.  The corresponding ground (Gross-Pitaevskii) eigenstate of the system has energy $E\approx9N_\mathrm{c}\hbar\omega_r$.  We analyze equilibrium configurations of this system with energies in the range $E\in[9.5,24.0]N_\mathrm{c}\hbar\omega_r$.   We note that our choice of parameters makes no reference to the physical number of atoms in an experimental system.  The classical-field approximation becomes asymptotically exact in the "classical" limit $N_\mathrm{c}\to\infty,\; U_0\to 0$, with $C_\mathrm{nl}$ fixed~\cite{Blakie08}, and our results correspond to this idealised classical-limit system.  Nevertheless, for realistic experimental systems in moderate and high-temperature regimes, the magnitude of quantum fluctuations is small compared to that of thermal fluctuations, and the PGPE provides a good description of the thermal Bose gas~\cite{Davis05,Simula06,Simula08b,Bezett08,Bezett09a,Bisset09}.     
\section{Correlation functions}\label{sec:Correlation_functions}
\subsection{Microcanonical formalism}
A central feature of the PGPE formalism is the microcanonical (ergodic) interpretation of field trajectories at equilibrium.  It is well known that the solutions of the nonlinear Schr\"odinger (Gross-Pitaevskii) equation exhibit \emph{stochastization}, leading to the ergodic equilibration of the system and the equipartition of the system energy~\cite{Damle96,Goral01,Berloff02,Lobo04,Connaughton05,Nunnenkamp07,Sinatra07}.  This property is also shared by trajectories of the PGPE for homogeneous~\cite{Davis01,Davis02,Davis03} and harmonically trapped~\cite{Blakie05,Davis05,Blakie08a,Wright08} systems.   In the ergodic interpretation, the PGPE trajectories provide a sampling of the microcanonical density 
\begin{equation}\label{eq:mu_density}
	P[\psi;E] = \left\{
	\begin{array}{rl}
		\mathrm{const} & \quad\quad H_\mathrm{CF}[\psi] = E \\
		0 & \quad\quad H_\mathrm{CF}[\psi] \neq E ,  
	\end{array} \right.  
\end{equation}
defined by the conserved first integrals of the system.  As the trajectories of $\psi(\mathbf{x},t)$ cover the density $P[\psi; E]$ densely, averages in the microcanonical density can be approximated by time averages along the trajectories of $\psi(\mathbf{x},t)$.  We therefore define correlation functions of the classical field as averages (or expectation values) of functionals of the field $\psi(\mathbf{x},t)$ in the density~Eq.~\reff{eq:mu_density}, which we approximate by time averages of the field (denoted by $\langle\cdots\rangle$).  Applying the microcanonical thermodynamic formalism of Rugh~\cite{Rugh97} to the PGPE trajectories, one finds that the PGPE system evolves over time to an equilibrium characterized by a well-defined chemical potential and temperature~\cite{Davis03,Davis05}, providing strong support for this ergodic interpretation of the classical-field dynamics.

A correlation function of particular importance in the application of the PGPE to partially condensed Bose systems is the covariance matrix
\begin{equation}\label{eq:obdm}
    G(\mathbf{x},\mathbf{x}') = \langle\psi^*(\mathbf{x})\psi(\mathbf{x}')\rangle,
\end{equation}
which is the classical-field analog of the quantum one-body density matrix~\cite{Goral02,Blakie05}.  As $G(\mathbf{x},\mathbf{x}')$ is Hermitian, we can diagonalize it to obtain a complete basis of eigenvectors $\{\chi_i(\mathbf{x})\}$ (the eigenmodes of the one-body density matrix) with real eigenvalues $\{n_i\}$ (the mean occupations of the eigenmodes at equilibrium), such that $G(\mathbf{x},\mathbf{x}') = \sum_i n_i \chi_i^*(\mathbf{x})\chi_i(\mathbf{x}')$.  By analogy to the Penrose-Onsager definition of condensation~\cite{Penrose56}, we identify the largest eigenvalue $n_0$ as the condensate occupation, and the corresponding eigenvector $\chi_0(\mathbf{x})$ as the condensate orbital.      
\subsection{U(1)-symmetric correlation functions}\label{subsec:Gauge_invariance}
An important feature of classical Hamiltonian systems is the relation between symmetries of the Hamiltonian and the conservation of quantities during the system's evolution (see, e.g., Refs.~\cite{Goldstein50,Sudarshan74}).  The Hamiltonian~\reff{eq:HCF} governing the classical-field dynamics is invariant under the $\mathrm{U(1)}$ (gauge) transformation $\psi(\mathbf{x})\to\psi(\mathbf{x})e^{i\theta}$, and this symmetry ensures that the normalization $N[\psi]=\int\!d\mathbf{x}\,|\psi(\mathbf{x})|^2$ of the classical field is conserved under the action of Eq.~\reff{eq:PGPE}.  Moreover, as Eq.~\reff{eq:HCF} has no explicit time dependence, the classical-field energy $H_\mathrm{CF}[\psi]$ is also conserved under the PGPE evolution.  We note that microcanonical density [Eq.~\reff{eq:mu_density}] of the PGPE system inherits the symmetries of the Hamiltonian~\reff{eq:HCF}; in particular, $P[\psi;E]$ is symmetric under the $\mathrm{U(1)}$ gauge transformation $\psi(\mathbf{x})\to\psi(\mathbf{x})e^{i\theta}$, and thus only the averages of quantities which are invariant under such transformations are nonzero in the microcanonical density. 

The $\mathrm{U(1)}$ phase symmetry of the classical-field Hamiltonian $H_\mathrm{CF}$ is of course also a symmetry of the ``fundamental" second-quantized Hamiltonian~\reff{eq:H_fundamental}, in which case it corresponds to the conservation of particle number under the action of the corresponding Heisenberg equation of motion for $\hat{\Psi}(\mathbf{x})$ \cite{Blaizot85}.  However, in traditional mean-field theories of Bose condensation, the condensate is assumed to acquire a definite phase, \emph{breaking} this symmetry~\cite{Anderson66,Leggett95}.  Theories built on this assumption do not strictly conserve particle number, and the grand-canonical formalism is typically employed to ensure conservation of the \emph{mean} particle number~\cite{Fetter99}.  Condensation in the field is then associated with the appearance of anomalous moments (moments of the field operator which are strictly zero in a state of fixed particle number) such as the condensate mean-field $\langle\hat{\Psi}\rangle$, and the anomalous thermal density $\langle \hat{\delta}\hat{\delta}\rangle$ (where $\hat{\delta}\equiv\hat{\Psi}-\langle\hat{\Psi}\rangle$).  In Ref.~\cite{Wright10b}, it was demonstrated that although the $\mathrm{U(1)}$ symmetry of the Hamiltonian $H_\mathrm{CF}$ is formally inherited by the microcanonical density~Eq.~\reff{eq:mu_density}, the classical-field solutions do exhibit a symmetry-breaking aspect, which allows for the calculation of anomalous averages such as $\langle\psi\rangle$ and $\langle\delta\delta\rangle$ in analogy to symmetry-breaking approximations to the second-quantized field theory.

An alternative approach to the theory of Bose condensates that respects the $\mathrm{U(1)}$ symmetry of the quantum-field Hamiltonian, and therefore conserves the number of atoms in the system, was presented by Girardeau and Arnowitt~\cite{Girardeau59}, and later rediscovered by Gardiner~\cite{Gardiner97a} and Castin and Dum~\cite{Castin98}.  This approach is distinguished from the symmetry-breaking approaches in that no spurious phase is assumed for the condensate,  and the fluctuations of the quantum field about the condensed mode are described in terms of modified ladder operators $\hat{b}_i \approx [\hat{a}_0^\dagger/(\hat{a}_0^\dagger\hat{a}_0^{\phantom{\dagger}})^{1/2}]\hat{a}_i$ (and their Hermitian conjugates) \cite{NoteB}, which demote a field quantum from a noncondensate mode $\phi_i(\mathbf{x})$ into the condensate mode $\phi_0(\mathbf{x})$ (and \emph{vice versa}).  In effect, these operators serve to describe the excitations of the system in a picture in which the (indeterminate) phase of the condensate mode is canceled, providing a rigorous basis for the construction of a theory of the fluctuations around the condensate.  Such an approach has been used by several authors in developing theories of Bose-Einstein condensation at zero and finite temperature \cite{Gardiner97a, Castin98, Morgan00, Gardiner07}.  

In this article, we take an analogous approach to characterizing fluctuations of the classical field about the condensate mode.  We introduce the fluctuation (or noncondensate) field  
\begin{equation}\label{eq:Lambda_defn}
    \Lambda(\mathbf{x},t) \equiv \frac{\alpha_0^*(t)}{\sqrt{\alpha_0^*(t)\alpha_0(t)}}\,\delta \psi(\mathbf{x},t),
\end{equation} 
where $\alpha_0(t) = \int\! d\mathbf{x}\, \chi_0^*(\mathbf{x})\psi(\mathbf{x},t)$ is the classical-field amplitude corresponding to the condensate mode $\chi_0(\mathbf{x})$, and
\begin{equation}
    \delta \psi(\mathbf{x},t) = \psi(\mathbf{x},t) - \chi_0(\mathbf{x})\!\int\!d\mathbf{x}'\,\chi_0^*(\mathbf{x}')\psi(\mathbf{x}',t),
\end{equation}
is the component of the classical field orthogonal to the condensate mode.  In terms of the Poisson brackets~\cite{Goldstein50,Sudarshan74} defined as
\begin{equation}
	\{F,G\}  \equiv \int\!d\mathbf{x}\;\left[\frac{\bar{\delta}F}{\bar{\delta}\psi(\mathbf{x})}\frac{\bar{\delta}G}{\bar{\delta}\psi^*(\mathbf{x})} - \frac{\bar{\delta}F}{\bar{\delta}\psi^*(\mathbf{x})}\frac{\bar{\delta}G}{\bar{\delta}\psi(\mathbf{x})}\right],  
\end{equation}
where the projected functional derivative operator $\bar{\delta}/\bar{\delta}\psi(\mathbf{x})=\sum_{n\in\mathbf{C}}Y_n^*(\mathbf{x})\partial/\partial \alpha_n$ \cite{Blakie08}, we easily find
\begin{equation}
   \Big\{\Lambda(\mathbf{x}),N[\psi]\Big\} = 0.
\end{equation}
The field $\Lambda(\mathbf{x})$ is therefore formally invariant under global rotations of the classical-field phase.  Thus, whereas moments of the noncondensate field which are not invariant under such rotations (e.g., $\langle \delta\psi\delta\psi\rangle$) necessarily vanish in the microcanonical ensemble due to the $\mathrm{U(1)}$ phase symmetry, the analogous moments of $\Lambda(\mathbf{x})$ (e.g., $\langle\Lambda\Lambda\rangle$), which we will refer to as \emph{anomalous} moments  of $\Lambda(\mathbf{x})$ in the remainder of this article, may legitimately acquire nonzero values in the ensemble.  
In the limit of a perfectly coherent condensate, for which $\langle|\alpha_0|^{2n}\rangle\equiv\langle|\alpha_0|^2\rangle^n$, it is clear that the first moment $\langle \Lambda \rangle = \langle\alpha_0^*\delta\psi\rangle/\sqrt{n_0}=0$, as is appropriate for a fluctuation variable.  It should be noted that more generally the vanishing of $\langle\Lambda(\mathbf{x})\rangle$ is not guaranteed \emph{a priori}~\cite{Gardiner07}; nevertheless we find $\langle\Lambda\rangle=0$ to good accuracy in our simulations.  
We note that normal correlation functions [those which are not anomalous; e.g., $\langle|\Lambda(\mathbf{x})|^2\rangle$] are \emph{manifestly} real, i.e., it is clear \emph{a priori} that complex conjugation has no effect on such correlation functions.  Anomalous correlation functions, by contrast, are not fundamentally constrained to be real in this manner.  However, we find that the anomalous correlation functions we consider in this article are predominantly real quantities (relative to a real condensate orbital), as a result of detailed balance in the equilibrium field.
\subsubsection{Numerical procedure}\label{subsubsec:U1_procedure}
Our procedure for calculating correlation functions is as follows: We form random initial states~\cite{Blakie05,Blakie08} with prescribed classical-field energies.  After evolving these initial states to equilibrium, we construct the one-body density matrix [Eq.~\reff{eq:obdm}] by ergodic averaging of the field trajectory.  The bulk features and simplest correlations of the field equilibrate on a time scale of $\sim 100\omega_r^{-1}$~\cite{Blakie08,Blakie08a,Wright10b}; however, we do not begin our ergodic averaging of the field trajectory until after an equilibration period of $1200\omega_r^{-1}$, to ensure that all moments of the field have settled down to their equilibrium values.  We then form 
ergodic averages from $1.8\times10^4$ equally spaced samples of the classical field taken over a subsequent period of $7200\omega_r^{-1}$ of the field evolution.  We diagonalize $G(\mathbf{x},\mathbf{x}')$ to find the condensate orbital $\chi_0(\mathbf{x})$ and its mean occupation $n_0$.  For the irrotational system we consider here, the condensate orbital has (aside from small numerical fluctuations) a uniform phase.  Before proceeding we absorb the overall phase of $\chi_0(\mathbf{x})$ into a global (time-independent) phase shift of the field trajectory used to construct the microcanonical ensemble.  This amounts to making a choice of gauge such that the condensate orbital is real and positive, and in terms of this orbital we define the condensate wave function $\Phi_0(\mathbf{x})\equiv\sqrt{n_0}\chi_0(\mathbf{x})$.  From each member of the ensemble we then form $\Lambda(\mathbf{x})$ according to Eq.~\reff{eq:Lambda_defn}, and we average products of $\Lambda(\mathbf{x})$ over the ensemble to form its moments.  We note that Eq.~\reff{eq:Lambda_defn} is singular when the overlap of the classical field $\psi(\mathbf{x},t)$ with the condensate orbital $\chi_0(\mathbf{x})$ vanishes.  However, the configurations of the field for which $\alpha_0=0$ are a set of measure zero in the microcanonical density, and in practical simulations we always find a finite value for $\alpha_0$, even in the high-temperature regime where the mode $\chi_0(\mathbf{x})$ is incoherent and undergoes large number fluctuations~\cite{Bezett09b}.
\section{Fluctuation correlations and interaction potentials}\label{sec:Correlations}
In this section we characterize the fluctuations of the noncondensate field by calculating the local second and third moments, and fourth cumulants~\cite{Gardiner04}, of the fluctuation field $\Lambda(\mathbf{x})$.  In terms of the $\mathrm{U(1)}$-symmetry preserving approach to calculating correlations we adopt in this article, we find that the field $\Lambda(\mathbf{x})$ exhibits fluctuations which are both anomalous (representing pairing effects induced by the condensate), and non-Gaussian (exhibiting deviations from the Gaussian ansatz for fluctuations assumed in the HFB theories~\cite{Blaizot85,Griffin96}).  These nontrivial correlations of the field can be related to corrections to the mean-field potentials experienced by the condensed and noncondensed atoms, due to many-body effects (see e.g., Refs.~\cite{Proukakis98a,Proukakis98b,Morgan00,Hutchinson00}).   

We focus here on a representative PGPE equilibrium, with energy $E=14.5N_\mathrm{c}\hbar\omega_r$, and corresponding \emph{condensate-band} condensate fraction $n_0/N_\mathrm{c}=0.50$.  This equilibrium is therefore a reasonably high-temperature state of the field, well above the validity regime of the simple Bogoliubov description \cite{Bogoliubov47,Fetter99,Castin98} of the noncondensate, while remaining far from the critical regime associated with the transition to the normal state \cite{Davis06,Bezett09b}.  
\subsection{Moments and cumulants of the noncondensate field}\label{subsec:Cumulants}
\subsubsection{Second moments}
There are two independent quadratic moments of the noncondensate field: the normal covariance matrix (or noncondensate density matrix) $\rho(\mathbf{x},\mathbf{x}') = \langle \Lambda^*(\mathbf{x})\Lambda(\mathbf{x}')\rangle$, and the anomalous covariance matrix (pair matrix) $\kappa(\mathbf{x},\mathbf{x}') = \langle \Lambda(\mathbf{x})\Lambda(\mathbf{x}')\rangle$.  All other quadratic moments can be related to these matrices by transposition and complex conjugation.  The spectral representation of the classical field allows us to calculate the full off-diagonal structure of these matrices, however, in this article we consider only their diagonal elements: the normal thermal density
\begin{equation}
    \rho(\mathbf{x}) = \langle \Lambda^*(\mathbf{x})\Lambda(\mathbf{x})\rangle,
\end{equation}
and the anomalous thermal density
\begin{equation}\label{eq:kappa_defn}
    \kappa(\mathbf{x}) = \langle \Lambda(\mathbf{x})\Lambda(\mathbf{x})\rangle.
\end{equation}
We note that in fact $\rho(\mathbf{x})\equiv\langle\delta\psi^*(\mathbf{x})\delta\psi(\mathbf{x})\rangle$, i.e., this quantity can be defined perfectly well without recourse to the considerations of Sec.~\ref{subsec:Gauge_invariance}.  By contrast, the anomalous density $\kappa(\mathbf{x})$ is distinct from the moment $\langle\delta\psi(\mathbf{x})\delta\psi(\mathbf{x})\rangle$, which of course vanishes in the microcanonical ensemble.  The physical origin of $\kappa(\mathbf{x})$ can be inferred by assuming that a factor $1/n_0$ can be factored out of the expectation value Eq.~\reff{eq:kappa_defn} [which should be a very good approximation for a coherent condensate mode $\chi_0(\mathbf{x})$], whereby we find that $\kappa(\mathbf{x})\sim\langle \alpha_0^*\alpha_0^*\delta\psi(\mathbf{x})\delta\psi(\mathbf{x})\rangle/n_0$.  The anomalous density thus quantifies the correlations of pairs of noncondensate atoms with pairs of condensate atoms due to the (classical wave-mixing analog of the) Bogoliubov pair-promotion process in which two condensate atoms scatter each other out of the condensate (and the time-reversed process), which is responsible for the well-known Bogoliubov particle-hole structure of excitations in the system~\cite{Fetter99,Castin98}.  The anomalous density can also be interpreted as a measure of the squeezing of the noncondensate field fluctuations~\cite{Cockburn10}.   

\begin{figure}
	\includegraphics[width=0.45\textwidth]{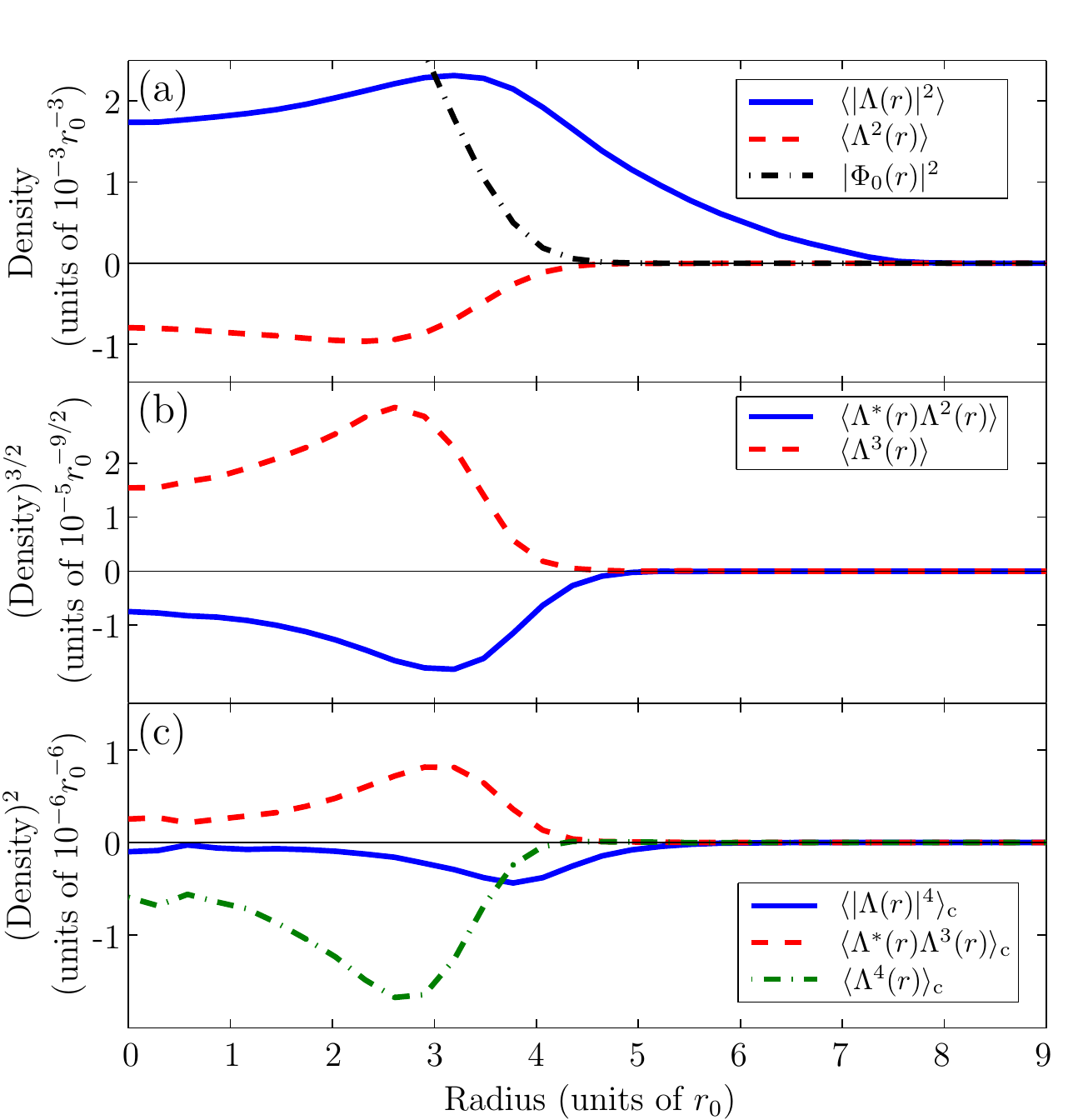}
    \caption{\label{fig:cumu_spatial} (Color online) Correlations of the fluctuation field $\Lambda(\mathbf{x})$ in a representative PGPE equilibrium with energy $E=14.5N_\mathrm{c}\hbar\omega_r$: (a)~Second moments of $\Lambda(\mathbf{x})$ (normal and anomalous thermal densities). (b) Third moments and (c) fourth cumulants of $\Lambda(\mathbf{x})$.} 
\end{figure}

In Fig.~\ref{fig:cumu_spatial}(a) we plot azimuthally averaged values of the thermal density $\rho(\mathbf{x})$ (solid line) and the anomalous thermal density $\kappa(\mathbf{x})$ (dashed line) on the $z=0$ plane.  
The condensate density $|\Phi_0(\mathbf{x})|^2$ (dot-dashed line) is plotted for comparison [the peak condensate density $|\Phi_0(0)|^2=12.7\times10^{-3} r_0^{-3}$].  The thermal density reaches its maximum in the periphery of the condensate, as is well known from mean-field theories (cf., for example, Ref.~\cite{Hutchinson97}).  The anomalous thermal density $\kappa(\mathbf{x})$ is real and negative to within statistical uncertainty due to the finite size of the ensemble \cite{Wright10b}, as expected, since in equilibrium the Bogoliubov pair-promotion process must be balanced by the corresponding time-reversed process.  We note that the anomalous density is almost entirely located within the extent of the condensate mode, and its magnitude exhibits a small dip around the trap center, in agreement with the results of previous studies \cite{Proukakis98b,Bergeman00,Hutchinson00,Wright10b}.
\subsubsection{Third moments}
We now turn our attention to the third moments of $\Lambda(\mathbf{x})$.  There are two independent third moments (or triplets~\cite{Laloe95,Proukakis96}) of $\Lambda(\mathbf{x})$.  In this article we only consider their diagonal elements, and we define
\begin{equation}\label{eq:lambda_defn}
    \lambda(\mathbf{x}) = \langle \Lambda^*(\mathbf{x})\Lambda(\mathbf{x})\Lambda(\mathbf{x})\rangle,
\end{equation}
and 
\begin{equation}
    \chi(\mathbf{x}) = \langle \Lambda(\mathbf{x})\Lambda(\mathbf{x})\Lambda(\mathbf{x})\rangle.
\end{equation}
In Fig.~\ref{fig:cumu_spatial}(b) we plot the functions $\lambda(\mathbf{x})$ (solid line) and $\chi(\mathbf{x})$ (dashed line), evaluated on the $z=0$ plane and azimuthally averaged.  Like the function $\kappa(\mathbf{x})$, both these moments are anomalous, but we find that they are real to within statistical uncertainty.  Both functions reside primarily in the central region where the condensate exists and, like $\rho(\mathbf{x})$ and $\kappa(\mathbf{x})$, each exhibits its maximal absolute value at some distance from the origin.  Although similar correlations have been discussed in the literature previously~\cite{Laloe95,Proukakis96,Proukakis98a,Hutchinson00,Kohler02}, we are not aware of any calculated forms for these functions with which to compare the results presented here.

We stress that the expectation values $\lambda(\mathbf{x})$ and $\chi(\mathbf{x})$ vanish implicitly in the self-consistent mean-field (HFB and HFB-Popov) theories, and their appearance here is a signature of the non-Gaussian nature of the fluctuation field $\Lambda(\mathbf{x})$.  The physical origin of the correlator $\lambda(\mathbf{x})$ can be inferred similarly to that of $\kappa(\mathbf{x})$: assuming a factor $1/\sqrt{n_0}$ can be factored out of the expectation value Eq.~\reff{eq:lambda_defn}, we find that $\lambda(\mathbf{x})\sim\langle \alpha_0^*\delta\psi^*\delta\psi\delta\psi\rangle/\sqrt{n_0}$.  The function $\lambda(\mathbf{x})$ is thus associated with processes in which two thermal atoms collide and one of them is scattered into the condensate (i.e., condensate \emph{growth} processes \cite{Zaremba99,Proukakis01,Gardiner02}).  In equilibrium this process is balanced by the time-reversed process, which is consistent with the real value of $\lambda(\mathbf{x})$ we find here, and represents the detailed balance of Beliaev and Landau processes (see, e.g., Ref.~\cite{Pethick02}) in the noncondensate, which we emphasize are not included in the HFB and HFB-Popov treatments~\cite{Morgan00}.  The correlation function $\chi(\mathbf{x})$ is less straightforward to interpret~\cite{NoteC}; however, analogous correlations are found to enter into the equations of motion for the pair matrix and the correlation function $\lambda(\mathbf{x})$ in generalized mean-field treatments of the finite-temperature Bose-gas dynamics \cite{Proukakis96,Proukakis01,Kohler02}.
\subsubsection{Fourth cumulants}
The appearance of the nonzero third moments $\lambda(\mathbf{x})$ and $\chi(\mathbf{x})$ shows that the fluctuations of the noncondensate field are not strictly Gaussian.  However, as the classical field we consider is weakly interacting, the fluctuations of $\Lambda(\mathbf{x})$ should be reasonably \emph{close} to Gaussian.  We therefore expect the fourth moments of the noncondensate field to be approximately equal to their naive Gaussian (Wick) factorizations \cite{Blaizot85} in terms of the second moments of $\Lambda(\mathbf{x})$; e.g.,  
\begin{equation}\label{eq:Wick}
    \langle |\Lambda(\mathbf{x})|^4 \rangle \approx 2\langle|\Lambda(\mathbf{x})|^2\rangle^2 + |\langle\Lambda^2(\mathbf{x})\rangle|^2.
\end{equation}
The differences between the actual fourth moments and their approximate Gaussian factorizations are quantified by the fourth cumulants of the field \cite{NoteD}.  There are three independent fourth cumulants of $\Lambda(\mathbf{x})$, corresponding to the three independent fourth moments of $\Lambda(\mathbf{x})$.  We consider here only the diagonal elements of these cumulants:  
\begin{eqnarray}\label{eq:cumu_def1}
    \langle |\Lambda(\mathbf{x})|^4\rangle_\mathrm{c} &=& \langle |\Lambda(\mathbf{x})|^4 \rangle - 2\langle |\Lambda(\mathbf{x})|^2\rangle^2 - |\langle \Lambda^2(\mathbf{x})\rangle|^2, \\ \label{eq:cumu_def2} 
    \langle \Lambda^*(\mathbf{x})\Lambda^3(\mathbf{x}) \rangle_\mathrm{c} &=& \langle \Lambda^*(\mathbf{x})\Lambda^3(\mathbf{x})\rangle - 3\langle |\Lambda(\mathbf{x})|^2\rangle\langle\Lambda^2(\mathbf{x})\rangle, \\ \label{eq:cumu_def3} 
    \langle \Lambda^4(\mathbf{x})\rangle_\mathrm{c} &=& \langle \Lambda^4(\mathbf{x})\rangle - 3 \langle\Lambda^2(\mathbf{x})\rangle^2. 
\end{eqnarray}
In Fig.~\ref{fig:cumu_spatial}(c) we plot the azimuthally averaged values of $\langle |\Lambda(\mathbf{x})|^4\rangle_\mathrm{c}$ (solid line), $\langle \Lambda^*(\mathbf{x})\Lambda^3(\mathbf{x})\rangle_\mathrm{c}$ (dashed line), and $\langle \Lambda^4(\mathbf{x})\rangle_\mathrm{c}$ (dot-dashed line) on the $z=0$ plane.  We note that these cumulants are small compared to the corresponding fourth moments.  For example, from Fig.~\ref{fig:cumu_spatial}(a) it can be inferred that the maximal value of the moment $\langle|\Lambda(\mathbf{x})|^4\rangle$ is $\sim10^{-5} r_0^{-6}$; i.e., $\sim10$ times the maximal value of $\langle|\Lambda(\mathbf{x})|^4\rangle_\mathrm{c}$.  We note that the fourth cumulants are somewhat more ``noisy'' than the lower correlations of the field; it is of course intuitively clear that the statistical demands for the accurate evaluation of cumulants will increase with the order of the cumulant.

The cumulants presented here do, however, indicate that the fluctuations of $\Lambda(\mathbf{x})$ exhibit a clear deviation from Gaussianity.  Of particular interest is the cumulant $\langle|\Lambda(\mathbf{x})|^4\rangle_\mathrm{c}$, as this measures the deviation of the fourth moment $\langle|\Lambda(\mathbf{x})|^4\rangle$ from its Wick factorization [Eq.~\reff{eq:Wick}].  This factorization becomes exact in the limit that the noncondensate fluctuations are Gaussian, and motivates the approximate factorizations of field-operator products made in mean-field theories in order to reduce the second-quantized Hamiltonian to a (self-consistent) quadratic form~\cite{Griffin96}.  In fact, Morgan~\cite{Morgan00} has shown that the treatment of the Hamiltonian term quartic in field operators in his second-order self-consistent calculation is equivalent 
to the Wick factorization employed in the HFB theories.  Effects which lead to the appearance of a nonzero fourth cumulant $\langle |\Lambda(\mathbf{x})|^4\rangle_\mathrm{c}$ at equilibrium thus enter at a higher order of perturbation theory.  We note that the cumulant $\langle|\Lambda(\mathbf{x})|^4\rangle_\mathrm{c}$ found here is negative, which can be understood intuitively: the field admits non-Gaussianity of its fluctuations so as to lower the interaction energy associated with a product of four noncondensate operators below the level obtained in self-consistent mean-field approaches.
\subsection{Many-body interaction effects}
We now relate the anomalous and non-Gaussian correlations of the classical field discussed in Sec.~\ref{subsec:Cumulants} to many-body interaction effects in the field which provide corrections to the mean-field potentials experienced by the condensate and noncondensed atoms.
\subsubsection{Mean-field potentials}
\begin{figure}
	\includegraphics[width=0.45\textwidth]{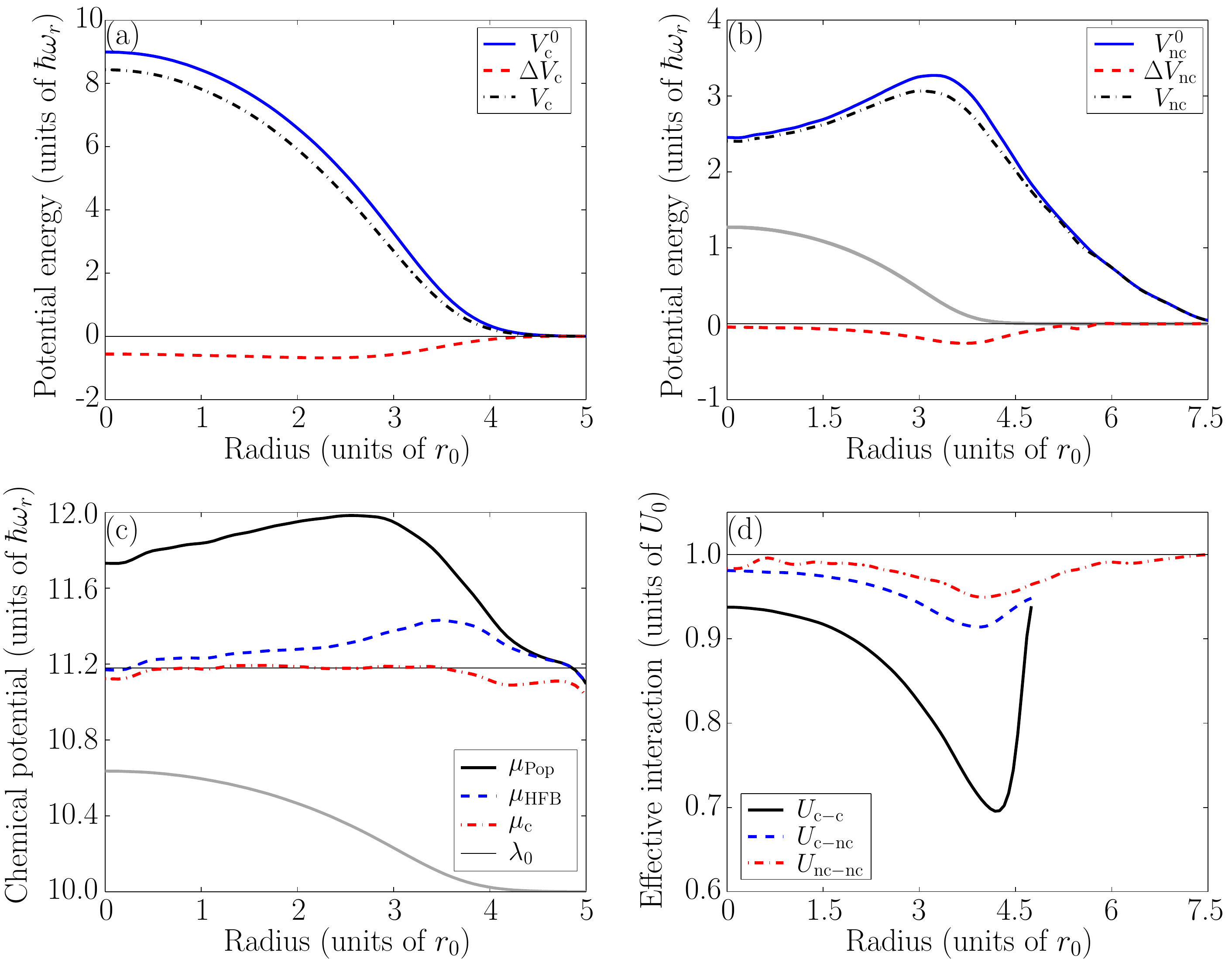}
    \caption{\label{fig:mf_potls} (Color online) Many-body interaction effects in the classical-field equilibrium:  (a-b) Mean-field potentials experienced by the condensate and their many-body corrections. (c) Local chemical potential of the classical-field condensate, in varying degrees of approximation (see text).  (d) Effective interaction strengths deduced from the correlations of $\Lambda(\mathbf{x})$.  Gray solid lines in (b) and (c) indicate the spatial density profile of the condensate.}
\end{figure}

It is well known that the appearance of the anomalous average in the generalized Gross-Pitaevskii equation of the HFB theory introduces many-body interaction effects which modify the collisions between condensate atoms \cite{Proukakis98a,Proukakis98b,Morgan00,Hutchinson00}.  The origin of this effect is the Bogoliubov pair-excitation process (Sec.~\ref{subsec:Cumulants}), which introduces the possibility that two colliding condensate atoms are both scattered into (possibly occupied) noncondensate modes, whereafter they may scatter off one another an arbitrary number of times before returning to the condensate.  In general, the entire series of such ``ladder diagram" processes \cite{Fetter71a} contributes to the effective interaction between condensate atoms in the finite-temperature system.  In the HFB theory~\cite{Griffin96}, it is found that the condensate $\Phi_0$ obeys a generalized Gross-Pitaevskii equation of the form $[H_\mathrm{sp}-\mu + U_0|\Phi_0|^2+2U_0\rho]\Phi_0 + U_0\kappa\Phi_0^* = 0$, and indeed this form is also obtained in the formalism of Ref.~\cite{Morgan00}.  The analyses of Refs.~\cite{Proukakis98a,Morgan00} show that the term $U_0\kappa\Phi_0^*$ corresponds to the introduction of the many-body $T$ matrix in the description of condensate-condensate interactions.  We therefore reformulate this term as an additional potential experienced by the condensate, which then obeys the nonlinear eigenvalue relation $\mathcal{L}_\mathrm{HFB}\Phi_0\equiv[H_\mathrm{sp}-\mu+U_0|\Phi_0|^2+2U_0\rho+U_0(\Phi_0^*)^2\kappa/|\Phi_0|^2]\Phi_0=0$.  The mean-field potential experienced by the condensate due to its own self-interaction 
\begin{equation}\label{eq:Vc0}
    V_\mathrm{c}^0(\mathbf{x})=U_0|\Phi_0(\mathbf{x})|^2,
\end{equation}
thus receives a correction~\cite{Proukakis98a,NoteE}
\begin{equation}\label{eq:dVc}
    \Delta V_\mathrm{c}(\mathbf{x})=\frac{U_0}{|\Phi_0(\mathbf{x})|^2}\mathrm{Re}\big\{[\Phi_0^*(\mathbf{x})]^2\langle\Lambda(\mathbf{x})\Lambda(\mathbf{x})\rangle\big\}.
\end{equation} 
In Fig.~\ref{fig:mf_potls}(a) we plot the bare mean-field potential $V_\mathrm{c}^0$ (blue/black solid line), the many-body correction $\Delta V_\mathrm{c}$ (dashed line), and their sum, the corrected condensate mean-field potential $V_\mathrm{c}=V_\mathrm{c}^0+\Delta V_\mathrm{c}$ (dot-dashed line).  
We indeed find that the presence of the anomalous average corresponds to a noticeable reduction of the mean-field potential $U_0|\Phi_0|^2$.  

Similarly to the mutual interaction of condensate atoms, the scattering of condensate atoms by noncondensate atoms is also affected by many-body processes in the finite-temperature Bose field:  Upon its interaction with a noncondensed atom, a condensate atom may be scattered out of the condensate, and undergo an arbitrary number of (repeated) interactions with the noncondensate atom before returning to the condensate.  The consideration of all such processes amounts to the inclusion of all ladder diagrams in the condensate-noncondensate interaction.   In an analysis beyond the usual mean-field (HFB) approach, Proukakis \emph{et al.} have shown~\cite{Proukakis98a} that the condensate obeys an equation of motion of the form $i\partial_t \Phi_0 = \mathcal{L}_\mathrm{HFB} \Phi_0 + U_0\lambda(\mathbf{x},t)$, where $\lambda(\mathbf{x},t)$ is the (time-dependent) triplet correlator corresponding to Eq.~\reff{eq:lambda_defn}. The adiabatic elimination of $\lambda(\mathbf{x},t)$ from this equation of motion leads to the introduction of the many-body $T$-matrix in the description of the condensate-noncondensate interactions, similar to the role of $\kappa(\mathbf{x})$ in the generalized GPE of the HFB theory.  The term $U_0\lambda$ can therefore be reformulated as an additional mean-field potential experienced by the condensate, yielding the equation of motion $i\partial_t\Phi_0=[\mathcal{L}_\mathrm{HFB}+U_0\Phi_0^*\lambda/|\Phi_0|^2]\Phi_0$.  The bare mean-field potential experienced by condensate atoms, due to the presence of the noncondensate,
\begin{equation}\label{eq:Vnc0}
    V_\mathrm{nc}^0(\mathbf{x}) = 2U_0\langle|\Lambda(\mathbf{x})|^2\rangle,
\end{equation}  
thus receives a many-body correction given by~\cite{Proukakis98a}
\begin{equation}\label{eq:dVnc}
    \Delta V_\mathrm{nc}(\mathbf{x}) = \frac{U_0}{|\Phi_0(\mathbf{x})|^2}\mathrm{Re}\big\{\Phi_0^*(\mathbf{x})\langle\Lambda^*(\mathbf{x})\Lambda(\mathbf{x})\Lambda(\mathbf{x})\rangle\big\}.
\end{equation}
It is important to note that while the correction Eq.~\reff{eq:dVc} arises due to the Bogoliubov processes which are accounted for to all orders by the Bogoliubov particle-hole structure of excitations in the HFB theories~\cite{Morgan00}, the correction of Eq.~\reff{eq:dVnc} corresponds to processes in which atoms are exchanged between the condensate and its excitations, which are not described in such theories.  In Fig.~\ref{fig:mf_potls}(b) we plot the bare potential $V_\mathrm{nc}^0$ (blue/black solid line), the correction $\Delta V_\mathrm{nc}$ (dashed line) and the corrected potential $V_\mathrm{nc}=V_\mathrm{nc}^0+\Delta V_\mathrm{nc}$ (dot-dashed line).  We find that the correction is localized near the boundary of the condensate, where the noncondensate density is maximal, however, it is nonzero at smaller radii also.  Again, we see the effect of the many-body correction is to weaken the mean-field potential experienced by the condensate.   
\subsubsection{Local chemical potential}
The properties of a superfluid are intimately connected to the \emph{phase} of the order parameter (i.e., the condensate), as the superfluid velocity is proportional to the gradient of the order-parameter phase: $\mathbf{v}_\mathrm{s}=(\hbar/m)\nabla \theta$ \cite{Anderson66,Pitaevskii80,Pitaevskii03}.  The chemical potential of the superfluid is, in general, a spatially varying quantity, corresponding to ($\hbar$ times) the phase-rotation frequency of the condensate~\cite{NoteF}.  A gradient in the chemical potential therefore corresponds to an acceleration of the superflow, and a stationary superfluid should exhibit a spatially uniform chemical potential~\cite{Anderson66}.  Here we define a local condensate chemical potential corresponding to the action of the aforementioned (generalized) mean-field operator $\mathcal{L}_\mathrm{HFB}+U_0\Phi_0^*\lambda/|\Phi_0|^2$ on the Penrose-Onsager condensate:
\begin{align}\label{eq:local_mu}
    \mu_\mathrm{c}(\mathbf{x}) =& \Big[H_\mathrm{sp}\Phi_0(\mathbf{x})\Big]/\Phi_0(\mathbf{x}) + U_0\Big[|\Phi_0(\mathbf{x})|^2 + 2\langle|\Lambda(\mathbf{x})|^2\rangle\Big] \nonumber \\
    &+\frac{U_0}{|\Phi_0(\mathbf{x})|^2}\mathrm{Re}\Big\{[\Phi_0^*(\mathbf{x})]^2\langle \Lambda^2(\mathbf{x})\rangle + \Phi_0^*(\mathbf{x})\langle \Lambda^*(\mathbf{x})\Lambda^2(\mathbf{x})\rangle\Big\}. 
\end{align}
We note that this local chemical potential is analogous to that derived by Zaremba~\emph{et al.}~\cite{Zaremba99} within a symmetry-breaking framework.  By evaluating $\mu_\mathrm{c}(\mathbf{x})$ and approximations to it that neglect the many-body correction terms appearing in Eq.~\reff{eq:local_mu}, we can assess the importance of these terms.  Specifically, in addition to the full form of $\mu_\mathrm{c}(\mathbf{x})$, we consider the chemical potential obtained by evaluating only those quantities on the first line of Eq.~\reff{eq:local_mu} [which we will refer to as the ``Popov" chemical potential, $\mu_\mathrm{Pop}(\mathbf{x})$], and that obtained by evaluating all but the last term of Eq.~\reff{eq:local_mu} [the ``HFB" chemical potential, $\mu_\mathrm{HFB}(\mathbf{x})$]. 

We calculate the action of $H_\mathrm{sp}$ on the condensate $\Phi_0$ in the spectral representation, and azimuthally average the resulting quantity  [$H_\mathrm{sp}\Phi_0(\mathbf{x})$], and the individual factors of the remaining terms on the RHS of Eq.~\reff{eq:local_mu}, on the $z=0$ plane in order to improve the estimates of these quantities and obtain radial representations of $\mu_\mathrm{c}(\mathbf{x})$, $\mu_\mathrm{Pop}(\mathbf{x})$, and $\mu_\mathrm{HFB}(\mathbf{x})$.  The results are presented in Fig.~\ref{fig:mf_potls}(c), along with the effective eigenvalue $\lambda_0$ of the condensate orbital, obtained from the temporal analysis of Ref.~\cite{Wright10b}.  We note first that the Popov-level chemical potential $\mu_\mathrm{Pop}(\mathbf{x})$ (thick solid black line) is somewhat nonuniform and overestimates the condensate eigenvalue $\lambda_0$ (thin solid black line) over the extent of the condensate mode [where the concept of $\mu_\mathrm{c}(\mathbf{x})$ is meaningful].  The HFB chemical potential $\mu_\mathrm{HFB}(\mathbf{x})$ (dashed line) is smaller than $\mu_\mathrm{Pop}(\mathbf{x})$ throughout the extent of the condensate mode, and approaches $\lambda_0$ much more closely at the smallest radii, while increasingly overestimating $\lambda_0$ at increasing radii~\cite{NoteG}.

By contrast, the full chemical potential $\mu_\mathrm{c}(\mathbf{x})$ (dot-dashed line) is much more uniform and in good agreement with the eigenvalue $\lambda_0$ aside from small discrepancies at the inner and outer extremes of the condensate orbital.  This result shows that the condensate orbital obtained from the Penrose-Onsager analysis is consistent with a mean-field picture of the condensate and the complementary thermal component of the field, \emph{provided} that the corrections to the mean-field potentials due to many-body effects are taken into account.  This explains the essentially uniform phase rotation of the (Penrose-Onsager) condensate mode discussed in Ref.~\cite{Wright10b}, and moreover demonstrates the consistency of the equilibrium classical-field formalism with other formulations of the finite-temperature Bose-gas problem (e.g., Ref.~\cite{Zaremba99}).  We stress that the quantities $\mu_\mathrm{Pop}(\mathbf{x})$ and $\mu_\mathrm{HFB}(\mathbf{x})$ do not correspond to the condensate chemical potentials that would be obtained in formal calculations using the corresponding self-consistent mean-field formalisms; such calculations would of course describe a condensate mode with a uniform chemical potential.  Nevertheless, the results presented here give an indication of the order of the error involved in calculations employing such approaches.
\subsubsection{Effective interactions}\label{subsubsec:effective_int}
We now reconsider the corrections to the mean-field potentials experienced by the condensate (and noncondensate) in terms of effective interaction strengths, which feature in the theory introduced in Refs.~\cite{Hutchinson98,Proukakis98b}, and reviewed in Ref.~\cite{Hutchinson00}.  Briefly, this so-called ``gapless-HFB" (GHFB) theory is based on the identification that the inconsistency of the HFB theory arises because the theory introduces many-body corrections to the condensate-condensate scattering, but that effects of this order (i.e., the many-body $T$-matrix approximation~\cite{Bijlsma97}) are not included in the description of interactions between the condensate and its excitations. The GHFB theory (of which there are two variants~\cite{Proukakis98b}) thus introduces spatially dependent \emph{effective} interaction strengths into the HFB-Popov theory, so as to approximate the effects of the many-body $T$-matrix corrections.   

In Fig.~\ref{fig:mf_potls}(d) we plot three effective interaction potentials deduced from the results of our classical-field simulation.  We define a condensate-condensate effective interaction by grouping the condensate-condensate mean-field potential $V_\mathrm{c}^0$ with its many-body correction $\Delta V_\mathrm{c}$, and absorbing the effect of the latter into a spatially dependent redefinition of coupling constant~\cite{Proukakis98b}:
\begin{equation}
    U_\mathrm{c-c}(\mathbf{x})\equiv U_0\Bigg(1+\frac{\langle \Lambda\Lambda \rangle}{\Phi_0^2}\Bigg).
\end{equation}
Making the substitution $U_0\to U_\mathrm{c-c}(\mathbf{x})$ in the bare condensate mean-field potential $V_\mathrm{c}^0$ [Eq.~\reff{eq:Vc0}] we obtain (by definition) the corrected potential $V_\mathrm{c}=V_\mathrm{c}^0+\Delta V_\mathrm{c}$.  The effective interaction $U_\mathrm{c-c}(\mathbf{x})$ plotted in Fig.~\ref{fig:mf_potls}(d) (solid line) thus provides an alternative visualization of the correction to the mean-field potential presented in Fig.~\ref{fig:mf_potls}(a), and is qualitatively consistent with the effective interactions presented in Refs.~\cite{Proukakis98b,Bergeman00,Hutchinson00}.

Next, we define an effective interaction appropriate to condensate-noncondensate scattering by absorbing the many-body correction $\Delta V_\mathrm{nc}$ to the mean-field potential $V_\mathrm{nc}^0$ experienced by the condensate, due to the presence of the noncondensate, into a local coupling constant 
\begin{equation}
    U_\mathrm{c-nc}(\mathbf{x})\equiv U_0\Bigg(1+\frac{\langle\Lambda^*\Lambda\Lambda\rangle}{2\Phi_0\langle\Lambda^*\Lambda\rangle}\Bigg).
\end{equation}
Making the substitution $U_0\to U_\mathrm{c-nc}(\mathbf{x})$ in $V_\mathrm{nc}^0$ [Eq.~\reff{eq:Vnc0}] clearly yields the corrected mean-field potential $V_\mathrm{nc}=V_\mathrm{nc}^0+\Delta V_\mathrm{nc}$.  We observe that $U_\mathrm{c-nc}$ [dashed line in Fig.~\ref{fig:mf_potls}(d)], like $U_\mathrm{c-c}$, is suppressed below the uniform value of the bare interaction potential $U_0$, and that this suppression is most pronounced around $r\approx4$ (i.e., the location of the edge of the condensate and the peak of the thermal-cloud density).  However, the suppression exhibited by $U_\mathrm{c-nc}$ is significantly less than that of $U_\mathrm{c-c}$, suggesting that the so-called \emph{G2} variant of the GHFB theory \cite{Proukakis98b,Hutchinson98,Hutchinson00}, in which $U_0$ is replaced by $U_\mathrm{c-c}$ everywhere it appears in the Gross-Pitaevskii and Bogoliubov-de Gennes equations, significantly overestimates the suppression of the noncondensate mean-field potential experienced by the condensate.  By contrast, the \emph{G1} variant of the theory retains the unmodified coupling constant $U_0$ in the mean-field potential $V_\mathrm{nc}$, and thus neglects any suppression of this potential. 

Finally in this section, we introduce an effective interaction strength for noncondensate-noncondensate interactions.  To define this quantity, we recall that the appearance of a negative fourth cumulant $\langle|\Lambda|^4\rangle_\mathrm{c}$ implies that the noncondensate-noncondensate interaction energy $E_4 = (U_0/2)\int\!d\mathbf{x}\,\langle |\Lambda|^4\rangle$ is suppressed somewhat below the value obtained by the Gaussian factorization of the moment $\langle |\Lambda|^4\rangle$ assumed in the HFB theory (Sec.~\ref{subsec:Cumulants}).  By assuming that the difference between $E_4$ and the corresponding approximate Gaussian expression can be accounted for by an effective (spatially-dependent) interaction strength associated with the mean-field potential due to noncondensate atoms, we identify  
\begin{equation}\label{eq:Uncnc}
    U_\mathrm{nc-nc}(\mathbf{x}) = U_0\Bigg(1+\frac{\langle|\Lambda|^4\rangle_\mathrm{c}}{2\langle|\Lambda|^2\rangle^2}\Bigg).  
\end{equation}
We observe that this quantity [dot-dashed line in Fig.~\ref{fig:mf_potls}(d)], like $U_\mathrm{c-c}$ and $U_\mathrm{c-nc}$, describes interactions which are suppressed most strongly around the condensate periphery. However, the maximal level of suppression here is of the order of $\sim5\%$, smaller again than that in the case of condensate-noncondensate scattering.  The small value of this correction is to be expected, as Eq.~\reff{eq:Uncnc} represents an average over all (coherent-region) excited-state collisions, many of which take place at high energies, where the many-body $T$~matrix reduces to the two-body $T$~matrix~\cite{Proukakis98b,Morgan00}.
\section{Dependence of correlations and interaction potentials on the field energy}\label{sec:E_dependence}
In this section, we consider the dependence of the noncondensate correlation functions and interaction potentials described in Sec.~\ref{sec:Correlations} on the total energy (and thus temperature) of the classical-field equilibrium. 
\subsection{Correlation functions}\label{subsec:E_dep_C_fns}
\begin{figure}
	\includegraphics[width=0.45\textwidth]{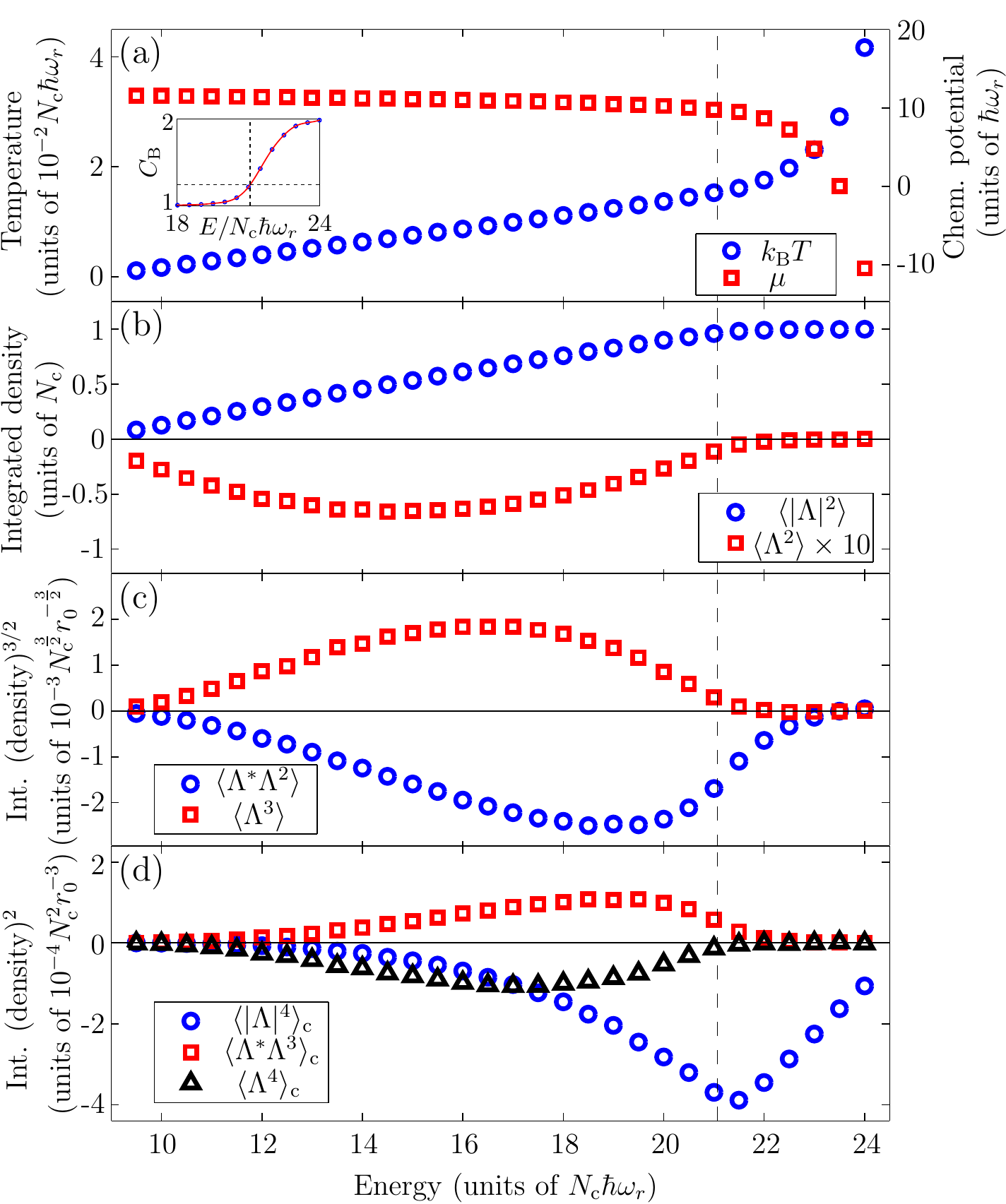}
    \caption{\label{fig:cumu_vs_E} (Color online) Dependence of thermodynamic parameters and noncondensate correlations on the classical-field energy. (a) Temperature and chemical potential of the field.  Magnitudes of spatially integrated (b) second moments, (c) third moments, and (d) fourth cumulants of $\Lambda(\mathbf{x})$.  Inset: Dependence of the Binder cumulant $C_\mathrm{B}$ on the field energy.  The solid line interpolates smoothly between the numerical data points (circles).  The dashed lines indicate the critical value $(C_\mathrm{B})_\mathrm{crit}=1.243$ and the corresponding critical field energy (see text).}
\end{figure}

In Fig.~\ref{fig:cumu_vs_E} we plot the spatially integrated values of correlation functions of the noncondensate field, obtained from PGPE equilibria with energies in the range $E\in[9.5,24.0]N_\mathrm{c}\hbar\omega_r$. 
The temperature and chemical potential of the field, obtained using the microcanonical thermodynamic formalism of Rugh~\cite{Rugh97,Davis03,Davis05}, are presented in Fig.~\ref{fig:cumu_vs_E}(a) for reference.  
In the inset to Fig.~\ref{fig:cumu_vs_E}(a), we plot the quantity $C_\mathrm{B}\equiv \langle|\alpha_0|^4\rangle/\langle|\alpha_0|^2\rangle^2$ as a function of the field energy.  Bezett and Blakie~\cite{Bezett09b} have suggested this quantity as the appropriate generalization of the \emph{Binder cumulant} associated with the phase transition of the homogeneous gas~\cite{Davis03} to the harmonically trapped case.  Campostrini \emph{et al.}~\cite{Campostrini01} found the critical value $(C_\mathrm{B})_\mathrm{crit}=1.243$ at the phase transition of the classical three-dimensional XY model, and this value was used to identify the critical temperature in PGPE simulations of the homogeneous Bose gas in Ref.~\cite{Davis03}.  The solid line in the inset to Fig.~\ref{fig:cumu_vs_E}(a) interpolates between the values of $C_\mathrm{B}$ obtained from our simulations (circles), and intersects the critical value $(C_\mathrm{B})_\mathrm{crit}$ (horizontal dashed line) at $E=21.1N_\mathrm{c}\hbar\omega_r$ (vertical dashed line), which we take as an estimate of the critical field energy for this system.  The critical energy is indicated by a vertical dashed line in Fig.~\ref{fig:cumu_vs_E}(a-d).  

Figure~\ref{fig:cumu_vs_E}(b) shows the spatially integrated normal thermal density $\int\!d\mathbf{x}\,\rho(\mathbf{x})$ (circles) and anomalous thermal density $\int\!d\mathbf{x}\,\kappa(\mathbf{x})$ (squares).  The results are similar to those obtained in \cite{Wright10b}, in which the anomalous correlation function $\kappa(\mathbf{x})$ was defined in terms of a symmetry-breaking interpretation of the classical-field trajectories.  It is important to note that the noncondensate population increases approximately linearly with energy (and thus temperature) up to the transition of the field to the noncondensed phase in the idealized ``PGPE system" consisting of a fixed field population distributed over the modally finite coherent region \cite{Connaughton05,Davis03,Davis05,Wright10b}.  Upon the consideration of the above-cutoff fraction of atoms, one obtains the expected, geometry-dependent scaling of the (non-)condensate fraction with temperature (see, for example, Ref.~\cite{Davis06}).  By contrast, the integrated anomalous density reaches its maximum absolute value at intermediate energies (temperatures), and vanishes as the system approaches the phase transition, in agreement with previous studies \cite{Proukakis98b,Bergeman00,Hutchinson00,Wright10b}.    

The integrated third moments $\int\!d\mathbf{x}\,\lambda(\mathbf{x})$ and $\int\!d\mathbf{x}\,\chi(\mathbf{x})$ are plotted in Fig.~\ref{fig:cumu_vs_E}(c).  The quantity $\int\!d\mathbf{x}\,\chi(\mathbf{x})$ reaches its maximum at intermediate energies, and vanishes along with $\int\!d\mathbf{x}\,\kappa(\mathbf{x})$ at $E\approx21.5N_\mathrm{c}\hbar\omega_r$.  The quantity $\int\!d\mathbf{x}\,\lambda(\mathbf{x})$, however, remains nonzero at the transition, and tends to zero at a somewhat higher energy.  We reiterate that although the definition [Eq.~\reff{eq:Lambda_defn}] of $\Lambda(\mathbf{x})$ is singular when $\alpha_0=0$, this condition does not occur in practical simulations~\cite{NoteH}.  Nonzero values for $\int\!d\mathbf{x}\,\lambda(\mathbf{x})$ are obtained at the phase transition as $\lambda(\mathbf{x})$ represents the scattering of atoms into and out of the ``condensate" mode $\chi_0(\mathbf{x})$, which persists in the critical regime, even though the mode $\chi_0(\mathbf{x})$ is no longer coherent.  At higher temperatures, where critical fluctuations subside and modes recover Gaussian fluctuations, $\lambda(\mathbf{x})$ vanishes. 

In Fig.~\ref{fig:cumu_vs_E}(d) we plot the integrated values of the fourth cumulants of the noncondensate field $\Lambda(\mathbf{x})$.  The cumulant $\langle \Lambda^4(\mathbf{x})\rangle_\mathrm{c}$ behaves similarly to $\kappa(\mathbf{x})$, as expected, given that this cumulant quantifies the difference between the purely anomalous (condensate-induced) moment $\langle \Lambda^4(\mathbf{x})\rangle$ and its naive factorization in terms of $\kappa(\mathbf{x})$ [see Eq.~\reff{eq:cumu_def3}].  Similarly to $\lambda(\mathbf{x})$, the integrated value of $\langle\Lambda^*\Lambda\Lambda\Lambda\rangle_\mathrm{c}$ vanishes at a slightly higher temperature.  From the definition [Eq.~\reff{eq:cumu_def2}] of this cumulant, it is clear that this behavior reflects that of the moment $\langle\Lambda^*\Lambda\Lambda\Lambda\rangle$, which can be understood in a similar manner to the behavior of $\lambda(\mathbf{x})$.  By contrast, the ``normal" cumulant $\langle|\Lambda|^4\rangle_\mathrm{c}$ reaches its \emph{maximum} absolute value around the transition to the normal phase.   This shows that many-body interaction effects are most important in the noncondensate field in this regime, in which the condensate is surrounded by a much larger region of the field which exhibits suppressed density fluctuations but no phase coherence (i.e., a \emph{quasicondensate}~\cite{Prokofev01}), as we discuss further in Sec.~\ref{sec:Suppression}.
\subsection{Interaction energies}
We now consider the variation of the many-body interaction effects identified in Sec.~\ref{subsubsec:effective_int} as functions of the field energy.  We introduce three interaction energies:
\begin{eqnarray}\label{eq:Ecc}
    E_\mathrm{c-c}\;\; &=& \frac{1}{2}\int\!d\mathbf{x}\,U_\mathrm{c-c}(\mathbf{x})|\Phi_0(\mathbf{x})|^4, \\ \label{eq:Ecnc}
   E_\mathrm{c-nc}\; &=& \phantom{\frac{1}{2}}\int\!d\mathbf{x}\,U_\mathrm{c-nc}(\mathbf{x}) |\Phi_0(\mathbf{x})|^2 \rho(\mathbf{x}), \\ \label{eq:Encnc}
    E_\mathrm{nc-nc} &=& \phantom{\frac{1}{2}}\int\!d\mathbf{x}\,U_\mathrm{nc-nc}(\mathbf{x}) \rho^2(\mathbf{x}),
\end{eqnarray}
corresponding to the condensate-condensate, condensate-noncondensate and noncondensate-noncondensate interactions, respectively.  We compare these three energies with the corresponding uncorrected energies, obtained from Eqs.~\reff{eq:Ecc}-\reff{eq:Encnc} by replacing the spatially dependent interaction strengths [$U_\mathrm{c-c}(\mathbf{x})$, etc] with the bare interaction strength $U_0$, which we denote by $E_\mathrm{c-c}^0$, etc.  

\begin{figure}
	\includegraphics[width=0.45\textwidth]{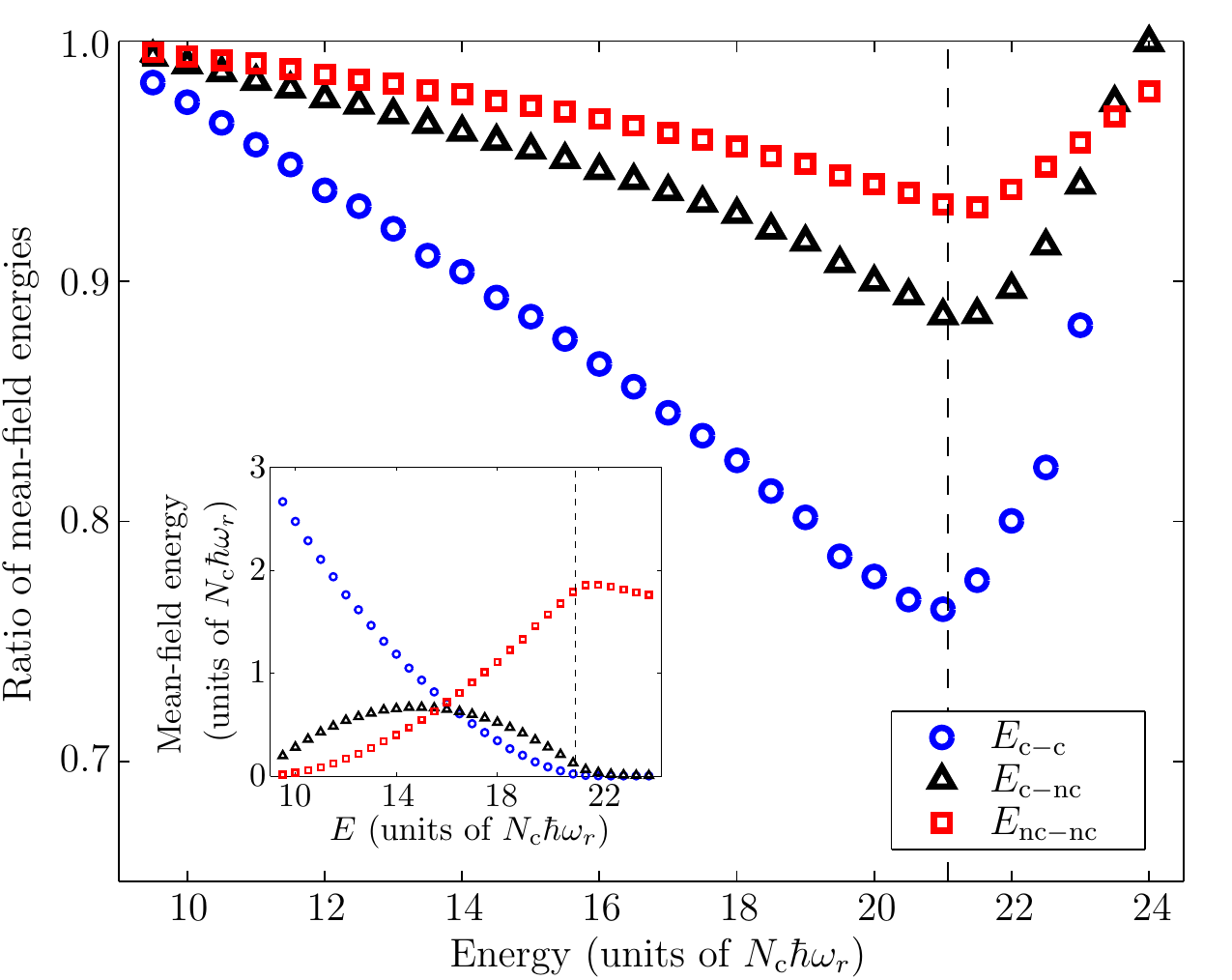}
    \caption{\label{fig:corrections_vs_E} (Color online) Dependence of the corrected interaction energies on the energy of the classical-field equilibrium.  Interaction energies are plotted as fractions of uncorrected interaction energies (see text).  The absolute values of the interaction energies are shown in the inset.}
\end{figure}

In Fig.~\ref{fig:corrections_vs_E} we plot the corrected interaction energies Eqs.~\reff{eq:Ecc}-\reff{eq:Encnc} as fractions of the corresponding uncorrected energies.  The approximate location of the phase transition determined from the analysis of the Binder cumulant $C_\mathrm{B}$ (Sec.~\ref{subsec:E_dep_C_fns}) is again indicated by a vertical dashed line. The absolute magnitudes of the corrected energies are shown in the inset to Fig.~\ref{fig:corrections_vs_E}.  
Considering the condensate-condensate interaction energy $E_\mathrm{c-c}$ (circles in Fig.~\ref{fig:corrections_vs_E}), we note that it continues to decrease (relative to $E_\mathrm{c-c}^0$) as $E$ increases towards the transition, and reaches its minimum at $E\approx21N_\mathrm{c}\hbar\omega_r$.  We contrast this with the behavior of $\int\!d\mathbf{x}\,\kappa(\mathbf{x})$ [Fig.~\ref{fig:cumu_vs_E}(b)], which reaches its maximum (absolute) value at $E\approx15N_\mathrm{c}\hbar\omega_r$.  At energies $E\gtrsim 23N_\mathrm{c}\hbar\omega_r$, where no condensate is present [i.e., the mode $\chi_0(\mathbf{x})$ is completely incoherent], the expression Eq.~\reff{eq:Ecc} is meaningless, and so we do not present data for $E_\mathrm{c-c}$ at these highest energies.  We find that the total condensate-condensate interaction energy is suppressed by $\sim 25\%$ at the transition, which is consistent with previous mean-field studies of harmonically trapped condensates~\cite{Proukakis98b,Bergeman00,Hutchinson00}, and should be contrasted with the complete vanishing of the condensate-condensate interaction at the transition temperature in the homogeneous case~\cite{Bijlsma97,Bijlsma96,Shi98}.   

Turning our attention to the energy $E_\mathrm{c-nc}$ of the condensate-noncondensate interaction (triangles in Fig.~\ref{fig:corrections_vs_E}), we observe that $E_\mathrm{c-nc}/E_\mathrm{c-nc}^0$, like $E_\mathrm{c-c}/E_\mathrm{c-c}^0$, reaches its minimum ($\sim0.88$) around the transition to the normal phase.  At higher energies (temperatures), the interaction returns to its bare value; in this limit, the ``condensate" mode $\chi_0(\mathbf{x})$ is simply a thermal (Gaussian) mode (i.e., $C_\mathrm{B}\approx2$), and Eq.~\reff{eq:Ecnc} reduces to the Hartree-Fock interaction energy between this mode and the rest of the field.  The energy of the noncondensate-noncondensate interaction (squares) is similarly suppressed most strongly at around $E\approx21N_\mathrm{c}\hbar\omega_r$, but is subject to less suppression than the other two interaction energies. 

We note the reasonable agreement between the critical energy estimated from the consideration of the Binder cumulant, and the energy at which maximal suppression of the field interactions occurs, which provides an independent estimate of the critical point (see Ref.~\cite{Proukakis98b} and references therein).  This suggests that the critical value $(C_\mathrm{B})_\mathrm{crit}=1.243$ appropriate to the phase transition of the homogeneous Bose gas does indeed yield a good estimate of the transition energy (temperature) in the harmonically trapped system. This is in contrast to the results of Ref.~\cite{Bezett09b} which suggested, based on an analysis of the field correlation length, that the critical value of $C_\mathrm{B}$ in the harmonically trapped system may be significantly lower than that in the homogeneous case. [Indeed the results presented in Figs.~\ref{fig:cumu_vs_E}~and~\ref{fig:corrections_vs_E} of this work suggest that $(C_\mathrm{B})_\mathrm{crit}$ slightly \emph{underestimates} the correct critical value].

Finally in this section, we note that the interaction energies considered here constitute an increasingly small fraction of the total field energy as the field energy (temperature) increases (inset to Fig.~\ref{fig:corrections_vs_E}).  In particular, in the critical region, the noncondensate-noncondensate interaction energy is $\lesssim 10\%$ of the total field energy, and the remainder is mostly comprised of the kinetic energy of the noncondensate fraction.  At the transition, the contribution of beyond-Gaussian fluctuations to the total field energy is therefore only of order $\sim1\%$ (cf. the agreement of HFB-Popov and PGPE calculations of the critical temperature to about this level reported in Ref.~\cite{Davis06}).
\section{Suppression of density fluctuations in the field}\label{sec:Suppression}
\begin{figure}
	\includegraphics[width=0.50\textwidth]{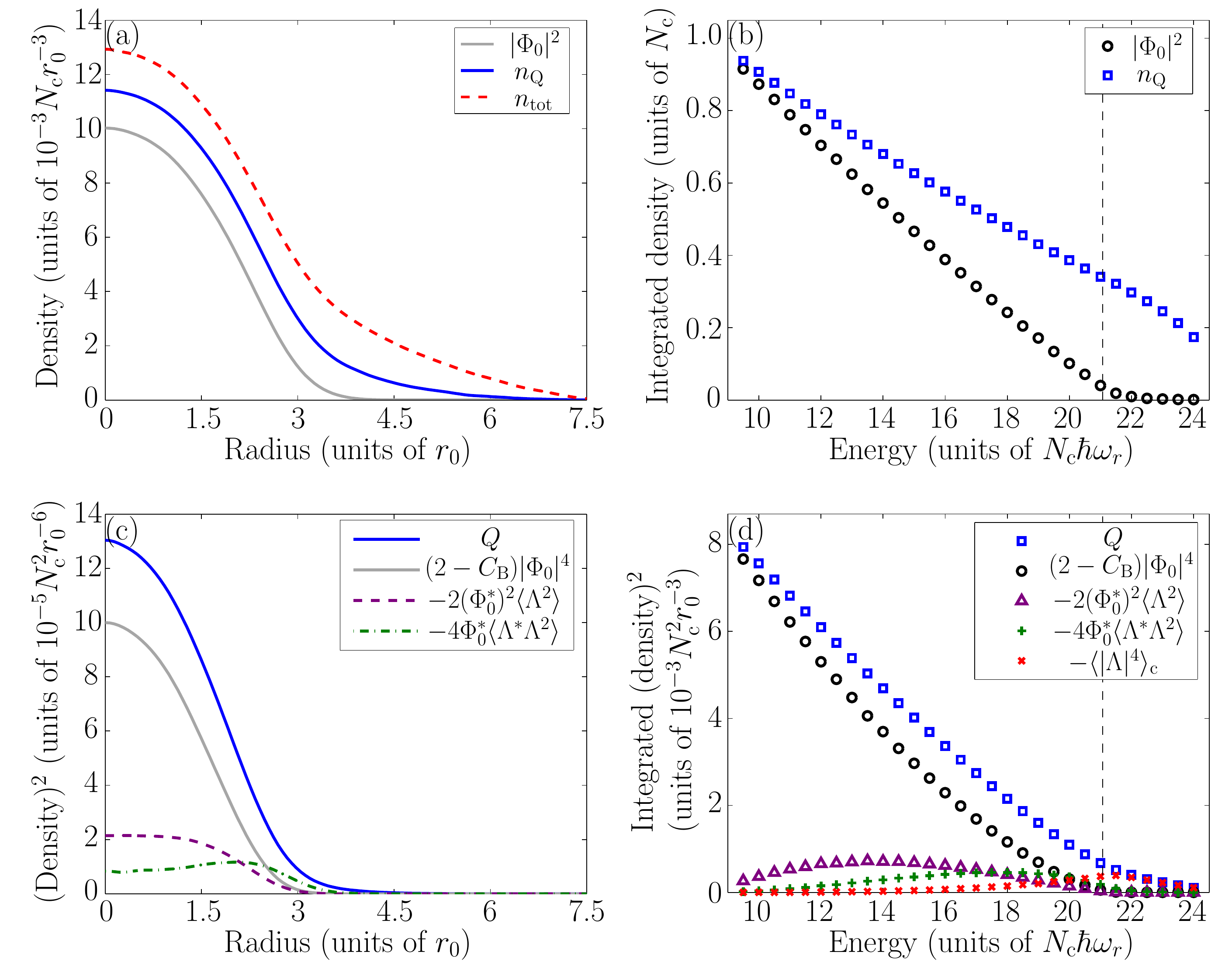}
    \caption{\label{fig:quasicond} (Color online) Suppression of density fluctuations in the classical field. (a) Condensate, quasicondensate, and total density profiles in a representative classical-field equilibrium with energy $E=18.0N_\mathrm{c}\hbar\omega_r$.  (b) Dependence of the total condensate and quasicondensate populations on the field energy. (c) Dominant contributions to the quasicondensation correlator $Q$ in the case $E=18.0N_\mathrm{c}\hbar\omega_r$.  (d) Dependence of the dominant contributions to $Q$ on the field energy.}
\end{figure}

We have seen in Sections~\ref{sec:Correlations}~and~\ref{sec:E_dependence} that the classical-field equilibria exhibit a negative fourth cumulant $\langle|\Lambda|^4\rangle_\mathrm{c}$, corresponding to the suppression of density fluctuations in the noncondensate field $\Lambda(\mathbf{x})$ below the Gaussian level assumed in the HFB and HFB-Popov formalisms.  We now consider the suppression of density fluctuations in the total classical field $\psi(\mathbf{x})$, and identify and compare the various contributions to this suppression.  We follow Ref.~\cite{Prokofev01} in considering the correlator 
\begin{equation}
    Q(\mathbf{x}) = 2\langle|\psi(\mathbf{x})|^2\rangle^2 - \langle|\psi(\mathbf{x})|^4\rangle,
\end{equation}
and defining the \emph{quasicondensate} density
\begin{equation}
    n_\mathrm{Q}(\mathbf{x})=\sqrt{Q(\mathbf{x})}=\Big[2-g^{(2)}(\mathbf{x})\Big]^\frac{1}{2}\langle|\psi(\mathbf{x})|^2\rangle,
\end{equation}
where the classical coherence function $g^{(n)}=\langle |\psi|^{2n}\rangle/\langle |\psi|^2\rangle^n$ \cite{Blakie05,Gardiner00,NoteI}.  The quasicondensate density $n_\mathrm{Q}(\mathbf{x})$ has been used by Bisset \emph{et al.}~\cite{Bisset09} to characterize the superfluid transition of the quasi-two-dimensional Bose gas (see also Ref.~\cite{Pietila10}).  We recall that the local correlation function $g^{(n)}$ adopts values of $g^{(n)}=n!$ and $g^{(n)}=1$ for ``normal" thermal fields (those without anomalous correlations) and purely coherent fields, respectively \cite{Glauber65,Blakie05}.  In Fig.~\ref{fig:quasicond}(a) we plot the (azimuthally averaged) quasicondensate density $n_\mathrm{Q}$ for the PGPE equilibrium with energy $E=18.0N_\mathrm{c}\hbar\omega_r$ (blue/black solid line), along with the condensate density $|\Phi_0|^2$ (gray solid line) and total field density $\langle|\psi|^2\rangle$ (dashed line).   
We observe that the quasicondensate density $n_\mathrm{Q}$ is everywhere greater than the condensate density, and in particular, $n_\mathrm{Q}$ remains significant at radii $r\gtrsim4r_0$, for which the condensate density essentially vanishes.  In Fig.~\ref{fig:quasicond}(b) we plot the integrated condensate density, i.e., the condensate population $n_0\equiv\int\!d\mathbf{x}\,|\Phi_0|^2$ (circles), and the integrated quasicondensate density $N_\mathrm{Q}\equiv\int\!d\mathbf{x}\,n_\mathrm{Q}$ (squares), as functions of the classical-field energy $E$.  We find that the quasicondensate population is greater than the condensate population for all energies considered, and that the difference between the two populations increases with increasing $E$.  Most notably, in the highest energy equilibria considered ($E\gtrsim22.0N_\mathrm{c}\hbar\omega_r$), in which no condensation remains, the quasicondensate population is still significant.  The presence of this quasicondensate at high energies (i.e., in the absence of condensation) clearly corresponds to the significant fourth cumulant $\langle|\Lambda|^4\rangle_\mathrm{c}$ found in this regime (Sec.~\ref{subsec:E_dep_C_fns}).

By expressing $\langle|\psi|^2\rangle$ and $\langle|\psi|^4\rangle$ in terms of $\alpha_0\chi_0$ and $\delta$, substituting $\delta\to\Lambda$ and assuming that the condensate can be factored out of these moments [e.g. $\langle|\alpha_0\chi_0|^2|\delta|^2\rangle\approx|\Phi_0|^2\langle|\Lambda|^2\rangle$ and $\langle (\alpha_0^*\chi_0^*)^2\delta^2\rangle\approx(\Phi_0^*)^2\langle\Lambda^2\rangle$], we obtain the approximate form \cite{NoteJ}
\begin{align}\label{eq:Q_approx}
    Q \approx (2-C_\mathrm{B})|\Phi_0|^4 &- \Bigg\{|\langle\Lambda\Lambda\rangle|^2 + 2\mathrm{Re}\big\{(\Phi_0^*)^2\langle\Lambda\Lambda\rangle\big\} \nonumber \\
    &\quad\;+4\mathrm{Re}\big\{\Phi_0^*\langle\Lambda^*\Lambda\Lambda\rangle\big\} + \langle|\Lambda|^4\rangle_\mathrm{c}\Bigg\}, 
\end{align}
where the Binder cumulant $C_\mathrm{B}$ accounts for the amplitude fluctuations of the condensate mode (see Sec.~\ref{subsec:E_dep_C_fns}).  We note that only by neglecting all but the first term on the RHS of Eq.~\reff{eq:Q_approx} do we obtain (in the limit $C_\mathrm{B}\to 1$) $n_\mathrm{Q}=|\Phi_0|^2$ (cf. Ref.~\cite{Dodd97b}).  The second term on the RHS of Eq.~\reff{eq:Q_approx} represents the \emph{enhancement} of density fluctuations in the noncondensate component relative to the normal Gaussian [$g^{(2)}=2$] level, due to the anomalous nature of its fluctuations (consider that $\langle|\Lambda|^4\rangle \approx 2\langle|\Lambda|^2\rangle^2+|\langle\Lambda^2\rangle|^2 > 2\langle|\Lambda|^2\rangle^2$) \cite{Gardiner01,Wright09a}, and thus subtracts from $Q$.  In practice we find that the contribution of this term is small.  The remaining terms on the RHS of Eq.~\reff{eq:Q_approx} all represent reductions in the density fluctuations of the field $\psi(\mathbf{x})$, and thus add to the quasicondensate density.

In Fig.~\ref{fig:quasicond}(c) we plot the (azimuthally averaged) quantity $Q$ for the case $E=18.0N_\mathrm{c}\hbar\omega_r$ (solid blue/black line), along with the three dominant contributions to this quantity: $(2-C_\mathrm{B})|\Phi_0|^4$ (solid gray line), $-2(\Phi_0^*)^2\langle\Lambda^2\rangle$ (dashed line) and $-4\Phi_0^*\langle\Lambda^*\Lambda\Lambda\rangle$ (dash-dot line).  The contributions of $|\langle\Lambda\Lambda\rangle|^2$ and $\langle|\Lambda|^4\rangle_\mathrm{c}$ are essentially negligible for this field energy, and the suppression of density fluctuations here is therefore primarily associated with the presence of a condensate in the field.  However, the extent to which density fluctuations are suppressed is somewhat greater than that due to the absence of density fluctuations in the condensate itself.  This suggests that an estimation of the condensate density based on an experimental measurement of $g^{(2)}$ could overestimate the size of the condensate considerably.  We plot the spatially integrated values of the correlator $Q$ and the dominant contributions to it as functions of the field energy in Fig.~\ref{fig:quasicond}(d).  We observe that the contribution of $-2(\Phi_0^*)^2\langle\Lambda\Lambda\rangle$ (triangles) is the dominant beyond-condensate contribution to $Q$ at the lowest energies, but reaches its maximum magnitude at $E\approx13.5N_\mathrm{c}\hbar\omega_r$, and is overtaken by the contribution of $-4\Phi_0^*\langle\Lambda^*\Lambda\Lambda\rangle$ (pluses) at higher energies.   This contribution is itself overtaken soon after by the contribution of the fourth cumulant $\langle|\Lambda|^4\rangle_\mathrm{c}$ (crosses), which is of course the dominant contribution in the critical regime.  

The results presented here should be compared with the behavior of the homogeneous Bose gas in two dimensions, for which it is found that the quasicondensate density is in general greater than the superfluid density, and that the quasicondensate correlations appear at temperatures well above the critical temperature for the appearance of superfluidity in the system~\cite{Kagan00,Prokofev01}.  Classical-field calculations for the harmonically trapped quasi-two-dimensional Bose gas appear to be consistent with a local-density version of this picture~\cite{Bisset09}.  It is perhaps not surprising that we observe similar behavior in the three-dimensional case considered here: the comparatively large transverse susceptibility of the field~\cite{Chaikin95} leads to the (local) destruction of order by phase fluctuations before size fluctuations become significant~\cite{Kleinert00}.  
\section{Conclusions}\label{sec:Conclusions}
We have presented a methodology for the calculation of general spatial correlations of the finite-temperature Bose gas, including anomalous correlations, in a classical-field approach.  We introduced $\mathrm{U(1)}$-symmetric classical-field variables, analogous to number-conserving quantum ladder operators considered by other authors~\cite{Girardeau59,Gardiner97a,Castin98}, in order to rigorously quantify the anomalous correlations in the microcanonical PGPE ensemble.  We have demonstrated that the finite-temperature Bose field exhibits non-Gaussian correlations which are classical in origin, and by relating these correlations to many-body corrections to the mean-field interaction potentials~\cite{Proukakis98a,Proukakis98b,Zaremba99,Morgan00}, we have explicitly demonstrated that the classical-field theory provides an intrinsic nonperturbative description of many-body processes in the field.  We discussed the role of these processes in determining the condensate mode, and demonstrated the consistency of the Penrose-Onsager~\cite{Penrose56} definition of condensation in the classical-field equilibrium.  Finally, we elucidated the contribution of the anomalous and non-Gaussian correlations of the noncondensate to the overall suppression of density fluctuations in the field, and discussed the distinction between the quasicondensate defined by the suppression of density fluctuations, and the true condensate defined by the suppression of phase fluctuations. 

Our demonstration of the intrinsic description of many-body interaction effects provided by the classical-field method underlines the utility of these techniques in describing low-dimensional systems~\cite{Bisset09,Foster10}, in which such effects can have profound consequences for the structure of the field equilibrium~\cite{Stoof93,Lee02,AlKhawaja02}.  Indeed our results illustrate the complexities introduced by the independence of phase and density fluctuations \cite{Kleinert00,Proukakis06} in the inhomogeneous finite-temperature Bose gas even in the comparatively straightforward three-dimensional case.  Moreover, although we have only considered systems at equilibrium, our results demonstrate that the classical-field model of the low-energy coherent region does describe the higher-order processes associated with quasiparticle damping, and the exchange of atoms between the condensate and its excitations, which are essential for the description of nonequilibrium dynamics of the finite-temperature Bose field~\cite{Proukakis98a,Fedichev98b,Giorgini00,Morgan03}.
\begin{acknowledgments}
While making revisions to this article, the authors were saddened to learn of the passing of Professor Allan Griffin, and would like to acknowledge discussions and correspondence with him that have helped to shape the ideas presented here.  N.P.P. would like to thank the University of Queensland node of the ARC Centre of Excellence for Quantum-Atom Optics for hospitality during his visit, and the UK EPSRC for financial support.  This work was supported by the Australian Research Council through the ARC Centre of Excellence for Quantum-Atom Optics (CE0348178). M.J.D. is the recipient of an ARC QEII fellowship (project DP1094025). 
\end{acknowledgments}
\bibliographystyle{prsty}

\begin{thebibliography}{100}

\bibitem{Anderson95}
{M. H. Anderson} {\it et~al.}, Science {\bf 269},  198  (1995).

\bibitem{Davis95}
K.~B. Davis {\it et~al.}, Phys. Rev. Lett. {\bf 75},  3969  (1995).

\bibitem{Bradley95}
C.~C. Bradley, C.~A. Sackett, J.~J. Tollett, and R.~G. Hulet, Phys. Rev. Lett.
  {\bf 75},  1687  (1995).

\bibitem{Shi98}
H. Shi and A. Griffin, Physics Reports {\bf 304},  1   (1998).

\bibitem{Dalfovo99}
F. Dalfovo, S. Giorgini, L.~P. Pitaevskii, and S. Stringari, Rev. Mod. Phys.
  {\bf 71},  463  (1999).

\bibitem{Proukakis08}
N.~P. Proukakis and B. Jackson, J. Phys. B {\bf 41},  203002  (2008). 

\bibitem{Bogoliubov47}
N.~N. Bogoliubov, J. Phys. (USSR) {\bf 11},  23  (1947).

\bibitem{Beliaev58}
S.~T. Beliaev, Sov. Phys. JETP {\bf 34},  289  (1958).

\bibitem{Griffin93}
A. Griffin, {\em Excitations in a Bose-Condensed Liquid} (Cambridge University
  Press, Cambridge, 1993).

\bibitem{Stoof99}
H.~T.~C. Stoof, J. Low Temp. Phys. {\bf 114},  11  (1999).

\bibitem{Rusch00}
M. Rusch, S.~A. Morgan, D.~A.~W. Hutchinson, and K. Burnett, Phys. Rev. Lett.
  {\bf 85},  4844  (2000).

\bibitem{Morgan03}
S.~A. Morgan, M. Rusch, D.~A.~W. Hutchinson, and K. Burnett, Phys. Rev. Lett.
  {\bf 91},  250403  (2003).

\bibitem{Morgan04}
S.~A. Morgan, Phys. Rev. A {\bf 69},  023609  (2004).

\bibitem{Griffin96}
A. Griffin, Phys. Rev. B {\bf 53},  9341  (1996).

\bibitem{Blaizot85}
J.-P. Blaizot and G. Ripka, {\em Quantum Theory of Finite Systems} (MIT Press,
  Cambridge, Massachusetts, 1985).

\bibitem{Proukakis96}
N.~P. Proukakis and K. Burnett, J. Res. Natl. Inst. Stand. Technol. {\bf 101},
  457  (1996).

\bibitem{Hugenholtz59}
N.~M. Hugenholtz and D. Pines, Phys. Rev. {\bf 116},  489  (1959).

\bibitem{Bijlsma97}
M. Bijlsma and H.~T.~C. Stoof, Phys. Rev. A {\bf 55},  498  (1997).

\bibitem{Shi94}
H. Shi, G. Verechaka, and A. Griffin, Phys. Rev. B {\bf 50},  1119  (1994).

\bibitem{Morgan00}
S.~A. Morgan, J. Phys. B {\bf 33},  3847  (2000).

\bibitem{Fedichev98b}
P.~O. Fedichev and G.~V. Shlyapnikov, Phys. Rev. A {\bf 58},  3146  (1998).

\bibitem{Proukakis98b}
N.~P. Proukakis, S.~A. Morgan, S. Choi, and K. Burnett, Phys. Rev. A {\bf 58},
  2435  (1998).

\bibitem{Hutchinson00}
D.~A.~W. Hutchinson {\it et~al.}, J. Phys. B {\bf 33},  3825  (2000). 

\bibitem{Stoof96}
H.~T.~C. Stoof, M. Bijlsma, and M. Houbiers, J. Res. Natl. Inst. Stand.
  Technol. {\bf 101},  443  (1996).

\bibitem{Proukakis98a}
N.~P. Proukakis, K. Burnett, and H.~T.~C. Stoof, Phys. Rev. A {\bf 57},  1230
  (1998).

\bibitem{Bijlsma96}
M. Bijlsma and H.~T.~C. Stoof, Phys. Rev. A {\bf 54},  5085  (1996).

\bibitem{Walser99}
R. Walser, J. Williams, J. Cooper, and M. Holland, Phys. Rev. A {\bf 59},  3878
   (1999).

\bibitem{Walser00}
R. Walser, J. Cooper, and M. Holland, Phys. Rev. A {\bf 63},  013607  (2000).

\bibitem{Giorgini00}
S. Giorgini, Phys. Rev. A {\bf 61},  063615  (2000).

\bibitem{Wachter01}
J. Wachter, R. Walser, J. Cooper, and M. Holland, Phys. Rev. A {\bf 64},
  053612  (2001).

\bibitem{Proukakis01}
N.~P. Proukakis, J. Phys. B {\bf 34},  4737  (2001). 

\bibitem{Imamovic01}
M. Imamovi\'c-Tomasovi\'c and A. Griffin, Journal of Low Temperature Physics
  {\bf 122},  617  (2001).

\bibitem{Blakie08}
P.~B. Blakie {\it et~al.}, Advances In Physics {\bf 57},  363  (2008).

\bibitem{Kagan97}
Y. Kagan and B.~V. Svistunov, Phys. Rev. Lett. {\bf 79},  3331  (1997).

\bibitem{Lobo04}
{C. Lobo}, {A. Sinatra}, and {Y. Castin}, Phys. Rev. Lett. {\bf 92},  020403
  (2004).

\bibitem{Brewczyk07}
M. Brewczyk, M. Gajda, and K. Rza\.zewski, J. Phys. B {\bf 40},  R1  (2007). 

\bibitem{Fetter71a}
A.~L. Fetter and J.~D. Walecka, {\em Quantum theory of many-particle systems}
  (McGraw-Hill, New York, 1971).

\bibitem{Baym99}
G. Baym, J.-P. Blaizot, M. Holzmann, F. Laloe, and D. Vautherin, Phys. Rev. Lett.
  {\bf 83},  1703  (1999).

\bibitem{Steel98}
M.~J. Steel {\it et~al.}, Phys. Rev. A {\bf 58},  4824  (1998).

\bibitem{Sinatra01}
A. Sinatra, C. Lobo, and Y. Castin, Phys. Rev. Lett. {\bf 87},  210404  (2001).

\bibitem{Polkovnikov03}
A. Polkovnikov, Phys. Rev. A {\bf 68},  053604  (2003).

\bibitem{Norrie05}
A.~A. Norrie, R.~J. Ballagh, and C.~W. Gardiner, Phys. Rev. Lett. {\bf 94},
  040401  (2005).

\bibitem{Davis01}
{M. J. Davis}, {S. A. Morgan}, and {K. Burnett}, Phys. Rev. Lett. {\bf 87},
  160402  (2001).

\bibitem{Blakie05}
{P.~B.~Blakie} and {M.~J.~Davis}, Phys. Rev. A {\bf 72},  063608  (2005).

\bibitem{Gardiner03}
{C.~W.~Gardiner} and {M.~J.~Davis}, J. Phys. B {\bf 36},  4731  (2003).

\bibitem{Davis02}
M.~J. Davis, S.~A. Morgan, and K. Burnett, Phys. Rev. A {\bf 66},  053618
  (2002).

\bibitem{Girardeau59}
M. Girardeau and R. Arnowitt, Phys. Rev. {\bf 113},  755  (1959).

\bibitem{Gardiner97a}
C.~W. Gardiner, Phys. Rev. A {\bf 56},  1414  (1997).

\bibitem{Castin98}
Y. Castin and R. Dum, Phys. Rev. A {\bf 57},  3008  (1998).

\bibitem{Girardeau98}
M.~D. Girardeau, Phys. Rev. A {\bf 58},  775  (1998).

\bibitem{NoteA}
{As noted in Ref.~\cite{Liu97}, due to the functional form of the Bose-gas
  equation of state, a ``nearly pure'' condensate with $\sim90\%$ condensate
  fraction corresponds to a temperature of order half the critical temperature.
  For a typical system population of $\sim10^5$ atoms, the lowest-lying
  excitations of the condensate, with energies on the order of the trap
  spacing~\cite{Stringari96b}, thus have thermal populations $\gtrsim 10$
  atoms, significantly larger than the zero-point ``vacuum'' mode occupation.
  As these modes are most strongly affected by the presence of the condensate,
  it is logical to expect that they form the dominant contribution to the
  non-trivial correlations of the noncondensed component of the field.}

\bibitem{Andersen04}
J.~O. Andersen, Rev. Mod. Phys. {\bf 76},  599  (2004).

\bibitem{Davis06}
M.~J. Davis and P.~B. Blakie, Phys. Rev. Lett. {\bf 96},  060404  (2006).

\bibitem{Wright10b}
T.~M. Wright, P.~B. Blakie, and R.~J. Ballagh, Phys. Rev. A {\bf 82},  013621
  (2010).

\bibitem{Davis05}
M.~J. Davis and P.~B. Blakie, J. Phys. A {\bf 38},  10259  (2005). 

\bibitem{Simula06}
{T. P. Simula} and {P. B. Blakie}, Phys. Rev. Lett. {\bf 96},  020404  (2006).

\bibitem{Simula08b}
T.~P. Simula, M.~J. Davis, and P.~B. Blakie, Phys. Rev. A {\bf 77},  023618
  (2008).

\bibitem{Bezett08}
A. Bezett, E. Toth, and P.~B. Blakie, Phys. Rev. A {\bf 77},  023602  (2008).

\bibitem{Bezett09a}
A. Bezett and P.~B. Blakie, Phys. Rev. A {\bf 79},  023602  (2009). 

\bibitem{Bisset09}
R.~N. Bisset, M.~J. Davis, T.~P. Simula, and P.~B. Blakie, Phys. Rev. A {\bf
  79},  033626  (2009).

\bibitem{Damle96}
K. Damle, S.~N. Majumdar, and S. Sachdev, Phys. Rev. A {\bf 54},  5037  (1996).

\bibitem{Goral01}
K. Goral, M. Gajda, and K. Rzazewski, Opt. Express {\bf 8},  92  (2001).

\bibitem{Berloff02}
N.~G. Berloff and B.~V. Svistunov, Phys. Rev. A {\bf 66},  013603  (2002).

\bibitem{Connaughton05}
C. Connaughton, C. Josserand, A. Picozzi, Y. Pomeau, and S. Rica, Phys. Rev. Lett.
  {\bf 95},  263901  (2005).

\bibitem{Nunnenkamp07}
A. Nunnenkamp, J.~N. Milstein, and K. Burnett, Phys. Rev. A {\bf 75},  033604
  (2007).

\bibitem{Sinatra07}
A. Sinatra, Y. Castin, and E. Witkowska, Phys. Rev. A {\bf 75},  033616  (2007). 

\bibitem{Davis03}
M.~J. Davis and S.~A. Morgan, Phys. Rev. A {\bf 68},  053615  (2003).

\bibitem{Blakie08a}
P.~B. Blakie, Phys. Rev. E {\bf 78},  026704  (2008).

\bibitem{Wright08}
T.~M. Wright, R.~J. Ballagh, A.~S. Bradley, P.~B. Blakie, and C.~W. Gardiner,
  Phys. Rev. A {\bf 78},  063601  (2008). 

\bibitem{Rugh97}
H.~H. Rugh, Phys. Rev. Lett. {\bf 78},  772  (1997).

\bibitem{Goral02}
K. G\'oral, M. Gajda, and K. Rz\c a\. zewski, Phys. Rev. A {\bf 66}, 051602  (2002).

\bibitem{Penrose56}
{O. Penrose} and {L. Onsager}, Phys. Rev. {\bf 104},  576  (1956).

\bibitem{Goldstein50}
H. Goldstein, {\em Classical Mechanics} (Addison-Wesley, Cambridge, MA, 1950).

\bibitem{Sudarshan74}
E.~C.~G. Sudarshan and N. Mukunda, {\em Classical Dynamics: A Modern
  Perspective} (Wiley, New York, 1974).

\bibitem{Anderson66}
P.~W. Anderson, Rev. Mod. Phys. {\bf 38},  298  (1966).

\bibitem{Leggett95}
A. Leggett,  in {\em Bose-Einstein Condensation in Atomic Gases}, edited by A.
  Griffin, D.~W. Snoke, and S. Stringari (Cambridge University Press,
  Cambridge, 1995).

\bibitem{Fetter99}
A.~L. Fetter,  in {\em Bose-Einstein Condensation in Atomic Gases}, edited by
  M. Inguscio, S. Stringari, and C.~E. Wieman (IOS Press, Amsterdam, 1999).

\bibitem{NoteB}
{A comparison of various possible definitions of number-conserving quantum
  fluctuation operators and their relative merits is given in
  Ref.~\cite{Gardiner07}.}

\bibitem{Gardiner07}
S.~A. Gardiner and S.~A. Morgan, Phys. Rev. A {\bf 75},  043621  (2007).

\bibitem{Bezett09b}
A. Bezett and P.~B. Blakie, Phys. Rev. A {\bf 79},  033611  (2009). 

\bibitem{Gardiner04}
C.~W. Gardiner, {\em Handbook of Stochastic Methods}, $3^\mathrm{rd}$ ed.
  (Springer-Verlag, Berlin, 2000).

\bibitem{Cockburn10}
S.~P. Cockburn, A. Negretti, N.~P. Proukakis, and C. Henkel, Phys. Rev. A {\bf
  83},  043619  (2011).

\bibitem{Hutchinson97}
D.~A.~W. Hutchinson, E. Zaremba, and A. Griffin, Phys. Rev. Lett. {\bf 78},
  1842  (1997).

\bibitem{Bergeman00}
T. Bergeman, D.~L. Feder, N.~L. Balazs, and B.~I. Schneider, Phys. Rev. A {\bf
  61},  063605  (2000).

\bibitem{Laloe95}
F. Lal\"oe,  in {\em Bose-Einstein Condensation in Atomic Gases}, edited by A.
  Griffin, D.~W. Snoke, and S. Stringari (Cambridge University Press,
  Cambridge, 1995).

\bibitem{Kohler02}
T. K\"ohler and K. Burnett, Phys. Rev. A {\bf 65},  033601  (2002).

\bibitem{Zaremba99}
E. Zaremba, T. Nikuni, and A. Griffin, J. Low Temp. Phys. {\bf 116},  277
  (1999).

\bibitem{Gardiner02}
{C.~W.~Gardiner}, {J.~R.~Anglin}, and {T.~I.~A.~Fudge}, J. Phys. B {\bf 35},
  1555  (2002).

\bibitem{Pethick02}
C.~J. Pethick and H. Smith, {\em Bose-Einstein Condensation in Dilute Gases}
  (Cambridge University Press, Cambridge, 2002).

\bibitem{NoteC}
{The function $\chi(\mathbf{x})$ represents correlations between three
  condensate particles and three noncondensate particles, which are induced by
  second-order processes which combine the Bogoliubov pair processes with the
  (Beliaev-Landau) condensate growth processes.}

\bibitem{NoteD}
{Note that as the first moments
  $\langle\Lambda\rangle=\langle\Lambda^*\rangle=0$, the second and third
  cumulants of $\Lambda(\mathbf{x})$ are simply the second and third moments
  themselves.}

\bibitem{NoteE}
{In Eqs.~\reff{eq:dVc},~\reff{eq:dVnc},~and~\reff{eq:local_mu}, we explicitly
  indicate the real part of the relevant products of $\Phi_0^*$ and (moments
  of) $\Lambda$. In the equilibrium situation we consider here, these products
  are in fact always real, regardless of the arbitrary choice of the phase of
  the condensate wave function $\Phi_0$ (Sec.~\ref{subsubsec:U1_procedure}).
  However, in nonequilibrium situations these products may have imaginary parts
  that correspond to sources for the condensate-density field
  $n_\mathrm{c}=|\Phi_0|^2$~\cite{Zaremba99}.}

\bibitem{Pitaevskii80}
L. Pitaevskii and E. Lifshitz, {\em Statistical Physics Part 2: Theory of the
  condensed state}, 2$^\mathrm{nd}$ ed. (Pergamon, Oxford, 1980).

\bibitem{Pitaevskii03}
L. {Pitaevskii} and S. Stringari, {\em Bose-{Einstein} Condensation} (Clarendon
  Press, Oxford, 2003).

\bibitem{NoteF}
{In fact, the condensate eigenvalue is in general distinct from the
  thermodynamic chemical potential of the system, differing by an additive
  factor which varies with the (inverse) number of condensate
  atoms~\cite{Proukakis98b, Morgan00, Bergeman00, Wright10b}.}

\bibitem{NoteG}
{Similar behavior has also been observed for the corresponding quantities in
  classical-field equilibria of the stochastic Gross-Pitaevskii
  equation~\cite{Cockburn10}.}

\bibitem{Hutchinson98}
D.~A.~W. Hutchinson, R.~J. Dodd, and K. Burnett, Phys. Rev. Lett. {\bf 81},
  2198  (1998).

\bibitem{Campostrini01}
M. Campostrini, M. Hasenbusch, A. Pelissetto, P. Rossi, and E. Vicari,
  Phys. Rev. B {\bf 63},  214503  (2001).

\bibitem{NoteH}
{Of course in the corresponding quantum-field development, the subspace of
  total condensate depletion must be handled carefully (see
  Ref.~\cite{Girardeau98}).}

\bibitem{Prokofev01}
N. Prokof'ev, O. Ruebenacker, and B. Svistunov, Phys. Rev. Lett. {\bf 87},
  270402  (2001).

\bibitem{Gardiner00}
C.~W. Gardiner and P. Zoller, {\em Quantum Noise}, $2^\mathrm{nd}$ ed.
  (Springer-Verlag, Berlin, 2000).

\bibitem{NoteI}
{Under reasonable assumptions regarding the correlations of the above-cutoff
  portion of the Bose field~\cite{Bezett08}, one finds that there is no
  contribution to $Q$ (or $n_\mathrm{Q}$) from the above-cutoff field component
  (which is not explicitly considered in this article).}

\bibitem{Pietila10}
V. Pietil\"a, T.~P. Simula, and M. M\"ott\"onen, Phys. Rev. A {\bf 81},  033616
   (2010).

\bibitem{Glauber65}
R.~J. Glauber,  in {\em Quantum Optics and Electronics}, edited by C. DeWitt,
  A. Blandin, and C. Cohen-Tannoudji (Gordon and Breach, New York, 1965),
  Chap.~Optical Coherence and Photon Statistics.

\bibitem{NoteJ}
{Note that the substitution $\delta \to \Lambda$ is in general an approximate
  one. We also neglect the term
  $\langle(\alpha_0^*\chi_0^*)^2\alpha_0\chi_0\delta\rangle$ and its conjugate
  as these correlations are strongly suppressed even in the inhomogeneous
  scenario considered here. [The first moment
  $\langle\Lambda(\mathbf{x})\rangle$ is closely related to this correlation
  function, and is similarly strongly suppressed; we find $\int\! d\mathbf{x}\,
  |\langle\Lambda(\mathbf{x})\rangle|^2 \lesssim 10^{-4}N_\mathrm{c}$ in all
  simulations we consider.]\phantom{.}}

\bibitem{Dodd97b}
R. Dodd, C. Clark, M. Edwards, and K. Burnett, Opt. Express {\bf 1},  284
  (1997).

\bibitem{Gardiner01}
C.~W. Gardiner and A.~S. Bradley, J. Phys. B {\bf 34},  4663  (2001). 

\bibitem{Wright09a}
T.~M. Wright, A.~S. Bradley, and R.~J. Ballagh, Phys Rev. A
  {\bf 80}, 053624 (2009). 

\bibitem{Kagan00}
Y. Kagan, V.~A. Kashurnikov, A.~V. Krasavin, N.~V. Prokofev, and B.~V. Svistunov,
  Phys. Rev. A {\bf 61},  043608  (2000).

\bibitem{Chaikin95}
P.~M. Chaikin and T.~C. Lubensky, {\em Principles of Condensed Matter Physics}
  (Cambridge University Press, Cambridge, 1995).

\bibitem{Kleinert00}
H. Kleinert, Phys. Rev. Lett. {\bf 84},  286  (2000).

\bibitem{Foster10}
C.~J. Foster, P.~B. Blakie, and M.~J. Davis, Phys. Rev. A {\bf 81},  023623
  (2010).

\bibitem{Stoof93}
H.~T.~C. Stoof and M. Bijlsma, Phys. Rev. E {\bf 47},  939  (1993).

\bibitem{Lee02}
M.~D. Lee, S.~A. Morgan, M.~J. Davis, and K. Burnett, Phys. Rev. A {\bf 65},
  043617  (2002).

\bibitem{AlKhawaja02}
U. Al~Khawaja, J.~O. Andersen, N.~P. Proukakis, and H.~T.~C. Stoof, Phys. Rev.
  A {\bf 66},  013615  (2002).

\bibitem{Proukakis06}
N.~P. Proukakis, Phys. Rev. A {\bf 73},  023605  (2006).

\bibitem{Liu97}
W.~V. Liu, Phys. Rev. Lett. {\bf 79},  4056  (1997).

\bibitem{Stringari96b}
S. Stringari, Phys. Rev. Lett. {\bf 77},  2360  (1996).

\end{thebibliography}

\end{document}